\pdfoutput=1
\documentclass{article}
\usepackage{arxiv}
\usepackage[utf8]{inputenc}
\usepackage{textcomp}
\usepackage{graphicx}
\usepackage{color}
\usepackage[version=4]{mhchem}
\usepackage{siunitx}
\usepackage{longtable,tabularx}
\usepackage{subcaption}
\usepackage{algorithm, algpseudocode}
\usepackage{booktabs}
\usepackage{multirow}
\usepackage{lipsum}
\usepackage{amssymb}
\usepackage{float}
\newtheorem{problem}{Problem}

\makeatletter
\newenvironment{breakablealgorithm}
{% \begin{breakablealgorithm}
		\begin{flushleft}
			\refstepcounter{algorithm}% New algorithm
			\hrule height.8pt depth0pt \kern2pt% \@fs@pre for \@fs@ruled
			\renewcommand{\caption}[2][\relax]{% Make a new \caption
				{\raggedright\textbf{\ALG@name~\thealgorithm} ##2\par}%
				\ifx\relax##1\relax % #1 is \relax
				\addcontentsline{loa}{algorithm}{\protect\numberline{\thealgorithm}##2}%
				\else % #1 is not \relax
				\addcontentsline{loa}{algorithm}{\protect\numberline{\thealgorithm}##1}%
				\fi
				\kern2pt\hrule\kern2pt
			}
		}{% \end{breakablealgorithm}
		\kern2pt\hrule\relax% \@fs@post for \@fs@ruled
	\end{flushleft}
}
\makeatother

\title{Domain-Knowledge-Aided Airborne Ground Moving Targets Tracking}
	
\author{
	Jianduo Chai \\
	School of Aerospace Engineering\\
	Beijing Institute of Technology\\
	Beijing, 100081 \\
	\texttt{3120210038@bit.edu.cn} \\
	\And
	Shaoming He\thanks{Corresponding Author.Email: 
		\textit{shaoming.he@bit.edu.cn}}  \\
	School of Aerospace Engineering\\
	Beijing Institute of Technology\\
	Beijing, 100081 \\
	\texttt{shaoming.he@bit.edu.cn} \\
	\And
	Hyo-Sang Shin \\
	School of Aerospace, Transport and Manufacturing\\
	Cranfield University\\
	Pittsburgh, 00000 \\
}

\begin{document}	
	\maketitle
	
	\begin{abstract}
		This paper investigates the problem of traffic surveillance using an unmanned aerial vehicle (UAV) and proposes a domain-knowledge-aided airborne ground moving targets tracking algorithm. To improve the accuracy of multiple targets tracking, the proposed algorithm incorporates domain knowledge into the joint probabilistic data association (JPDA) filter as state constraints. The domain knowledge considered in this paper includes both road information extracted from a given map and local traffic regulations. Conventional track update method is modified to enhance the capability of recognition of temporarily track loss. A variable structure multiple model (VS-MM) method is developed to assign the road segment to a given target. The effectiveness of proposed algorithm is demonstrated through extensive numerical simulations.
	\end{abstract}

	\keywords{Traffic surveillance, Ground targets tracking, UAV, Domain-knowledge-aided, Multiple constraints, JPDA}
	
	\section{Introduction}\label{sec1}
	
	Small-scale UAVs have been widely-used in many military and civilian applications in recent years, e.g., urban crowd surveillance \cite{n04,new01}, ground vehicles monitoring \cite{n03,RN34,RN31,RN35}, search and rescue \cite{n05,n06}, wild-land fire management \cite{n01}, and borders surveillance \cite{n07}. One of the key enablers to improve the level of autonomy for UAV is autonomous ground targets tracking, which aims to find the states of targets of interest from the time history of noisy onboard measurements. Notice that this is a constrained nonlinear multiple targets tracking problem, where the constraints are topological limits and local traffic regulations. Up to now, this problem has not been well solved and requires new advances on many techniques involving constrained nonlinear state estimation, data association, track confirmation and termination.
	
	In airborne target tracking, the sensor measurement is generally nonlinear with respect to the target state and many elegant nonlinear estimation filters are reported in the literature for this application. The authors in \cite{RN35,RN46} utilized an extended Kalman filter (EKF) to estimate the state of a single ground vehicle and the estimated information was used to analyze the behavior of the vehicle. A constraint particle filter was developed in \cite{RN33} to predict the target existence area, guaranteeing that the target of interest remains tracked by UAVs. Reference \cite{RN34} utilized Recursive Bayesian estimation (RBE) for target state estimation, and proposed a novel method to approximate the likelihood density corresponding to no detection cases. To improve the estimation performance with high nonlinearity, a particle filter (PF) was proposed in \cite{RN45} for ground moving targets tracking over a finite horizon. The authors \cite{RN36} and \cite{RN32} leveraged an interacting multiple model (IMM) filter \cite{RN11} as a baseline for maneuvering target tracking. Reference \cite{e3} integrated variable structure IMM with EKF for a specific application of tracking move-stop-move target with airborne ground moving target indicator. However, most previous studies of airborne ground target tracking generally focus on single target tracking approaches. Realistic scenarios are often encountered with the multiple ground moving targets tracking problem and this is much more challenging than a single target due to data association uncertainty, constraint handling, track confirmation, deletion and continuation \cite{RN31,RN30,e-mht,new03}.
	
	Technically speaking, the Multiple Hypothesis Tracking (MHT) algorithm, first proposed in \cite{RN38}, is believed to be the Bayesian optimal filter for multiple targets tracking \cite{RN39}. Although it is recognized that MHT can track multiple targets with high accuracy by a delayed decision logic\cite{new02,new04,new05}, the computational complexity of MHT prohibits its application in the small-scale UAVs with limited computational resource. Compared to the MHT, the joint Probabilistic Data Association (JPDA) algorithm provides satisfactory performance with reduced computational burden for multiple target tracking in a cluttered environment \cite{RN9,pda,he2020trajectory}. However, the original JPDA assumes that the number of targets is known in advance and constant across the tracking process. To relax this assumption, the authors in \cite{RN14,he2020information} leveraged a modified JPDA that utilizes probabilistic birth model for new born target confirmation and death target termination. However, applying existing JPDA or its variants in ground vehicle surveillance still face with two challenges. On one hand, the moving trajectory of a ground vehicle is naturally constrained by the topological pattern and local traffic regulations. Incorporating this domain knowledge as state constraint in JPDA is helpful to improve the estimation performance. Even though there are some studies about constraint filter design \cite{RN8,RN6,RN43,RN7,RN22,ckf}, these algorithms are only dedicated to single target tracking and the extension to multiple targets tracking requires careful adjustment. On the other hand, the conventional track update logic  \cite{RN14}, capable of recognizing targets that are covered by possible birth regions, fails to reconfirming temporarily disappeared targets, which are typical phenomena in our application scenarios due to occasional blocking by buildings or trees.
	
	In this paper, a novel domain-knowledge-aided airborne ground moving targets tracking algorithm is proposed to fill the gap identified above. We approximate each road at the surveillance region by multiple straight-line segments, and the geographic information (e.g., heading direction) of each segment is extracted as domain knowledge. A new JPDA variant that actively leverages the domain knowledge, i.e., topological information and local traffic regulations, as state constraints is developed to improve the estimation performance. A modified track update approach is then proposed to ensure that the reappeared targets on the midway of a road can be confirmed with proper constraints. Variable structure multiple model (VS-MM) algorithm is utilized for associating targets with correct segment constraint models. By considering that a vehicle moves on a specific road segment as a constraint hypothesis, the VS-MM concept is leveraged for hypothesis propagation and a hard decision logic is suggested for hypothesis determination. The proposed algorithm is evaluated with extensive numerical simulations with various different conditions.
	
	The main contributions of this paper are twofold. On one hand, we propose a complete VS-MM tracking algorithm for multiple ground vehicles tracking. The algorithm is capable of new born target initialization, data association, state estimation, dead track deletion, reappeared target recognition, constraint handling and mode propagation. On the other hand, external domain knowledge information is extracted, modeled and integrated into the JPDA filter for constrained multiple targets tracking. This is demonstrated to significantly improve the estimation performance by providing reliable and robust tracking quality across various different conditions. 
	
	The remaining paper is organized as follows. Section \ref{sec2} presents some backgrounds and preliminaries of this paper. Section \ref{sec3} extracts map information through a road approximation method. A modified JPDA filter which is integrated with multiple constraints and able to reconfirm temporarily disappeared targets is proposed in Section \ref{sec4}. Section \ref{sec5} develops an approach using VS-MM concept for constraint hypothesis determination. The numerical simulations are presented in Section \ref{sec6}. Finally, some conclusions are given in Section \ref{sec7}.
	
	%%%%%%%%%%%%%%%%%%%%%%%%%%%%%%%%%
	\section{Backgrounds and Preliminaries}\label{sec2}
\begin{figure*}[hbt!]
		\centering
		\includegraphics[width=0.7\linewidth]{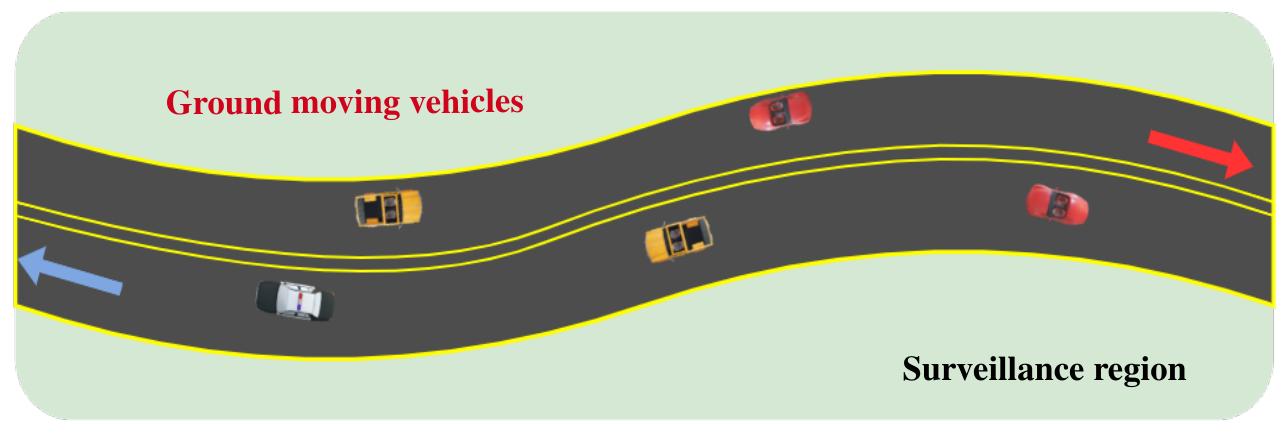}
		\caption{Scenario of aerial surveillance of multiple ground moving vehicles.}\label{Aerial surveillance of multiple ground moving vehicles}
	\end{figure*}

	This section first introduces the target dynamics model and sensor measurement model that will be used in the following sections. Then, we state the problem formulation and briefly introduce the logical flow of the proposed algorithm. We consider a scenario, as depicted in Fig. \ref{Aerial surveillance of multiple ground moving vehicles}, that involves one hovering UAV to monitor a certain surveillance region with no intersections and the ground vehicles are moving with varying speeds that are supposed to satisfy certain limits due to local traffic regulations. Additionally, the ground moving vehicles are also constrained on the road, i.e., their trajectories are constrained within the road boundaries. 
	
	\subsection{Target Model}
	Denote $\left\{\mathbf{x}\right\}_{k} \triangleq \left\{\mathbf{x}_{k}^{i} \vert i=1,\ldots,N_k\right\}$ as the collection of states of targets at time instant $t_k$ with $\mathbf{x}_{k}^{i}$ being the state vector of the $i$th target at time instant $t_k$ and $N_k$ being the unknown number of targets at time instant $t_k$. Notice that the maneuverability of the ground moving vehicle is usually limited and hence the moving speed can be assumed as piece-wise constant. For this reason, we leverage the constant velocity (CV) model to describe the motion of the vehicle., i.e., the state vector of the $i$th target at time instant $t_k$ can be defined by $\mathbf{x}^{i}_{k}=\left[p^i_{x,k},p^i_{y,k},p^i_{z,k}, v^i_{x,k},v^i_{y,k},v^i_{z,k}\right]^{\text{T}}$ with $\left[p^i_{x,k},p^i_{y,k},p^i_{z,k}\right]^{\text{T}}$ and $\left[v^i_{x,k},v^i_{y,k},v^i_{z,k}\right]^{\text{T}}$ being the 3-D position and velocity vectors, respectively. Then, the system dynamics can be modeled by a linear Markov chain as
	\begin{align}\label{dynamic model}
		\mathbf{x}_k^{i}=\mathbf{F}_{k}\mathbf{x}_{k-1}^{i}+\mathbf{v}_{k}^{i}
	\end{align}
	where $\mathbf{v}_{k}^{i}$ is white Gaussian noise with zero-mean and covariance matrix $\mathbf Q_{k}$. $\mathbf{F}_{k}$ is state transition matrix and is correlated with the sampling period $t_{k\vert k-1}\triangleq t_{k}-t_{k-1}$ as
	\[\mathbf{F}_k=
	\left[
	\begin{array}[array]{cccc}
		\mathbf{I}_{3}&t_{k\vert k-1}\mathbf{I}_{3}\\
		\mathbf{0}_3&\mathbf{I}_{3}
	\end{array}\right]
	\]
	where $\mathbf{I}_{3}$ denotes the $3\times 3$ identity matrix, and $\mathbf{0}_3$ stands for the $3\times 3$ zero matrix. 
	
	Since the ground moving vehicles are supposed to satisfy certain constraints, e.g., traffic regulations and local topography, which can be extracted and abstracted as multiple constraints of the targets as 
	\begin{align}
		g({\mathbf x}_{k}^{i})\preccurlyeq {\mathbf d}
		\label{constraint condition}
	\end{align}
	where $g(\cdot)$ represents the constraint function and $\mathbf d$ denotes the given upper bound.
	
	% \emph{Note that except for the inequality constraint conditions, the equality is also included in equation $(\ref{constraint condition})$.}
	
	\subsection{Sensor Model}
	We define $\mathbf Z^{k}\triangleq\left\{\mathbf Z_{1},\ldots,\mathbf Z_{k}\right\}$ as the collected measurement sequence up to time instant $t_k$ with $\mathbf Z_{k}\triangleq \{\mathbf{z}_{k}^j\vert j=0,1,\ldots,M_k\}$ being the received measurement set at time instant $t_k$. Since the sensor might be subject to some uncertainties, non-unity detection probability and external environmental disturbances, e.g., sunlight reflection or inference from other objects, the measurement might originate from a target of interest or be a false alarm. For simplicity, we use $j=0$ to represent the dummy measurement or false alarm, which is assumed to be averagely distributed over the surveillance region with intensity $\lambda_f$. The target-originated measurements are detected independently over time with known detection probability $P_D$ \cite{pda} can be modeled by 
	\begin{align}
		\mathbf{z}_{k}^j=h(\mathbf{x}_k^i)+\mathbf{w}_k^j\label{mea}
	\end{align}
	where $\mathbf{w}_k^j$ is zero-mean white Gaussian noise with covariance $\mathbf R_{k}$ and is assumed to be independent with the process noise $\mathbf{v}_{k}^i$. 
	
	In this study, the UAV, hovering at a known position $\left[p_{ x}^U, p_{ y}^U, p_{ z}^U\right]^T$, is assumed to be equipped with an active sensor that provides the measurements of the relative range $r_k^i$, elevation angle $\theta_k^i$ and azimuth angle $\xi_k^i$ between the UAV and the $i$th target. Therefore, the nonlinear function $h(\mathbf{x}_k^i)$ that models the relationship between the target state and the observed state, can be readily obtained in Eq. \eqref{mtss}.
	\begin{figure*}[hbt!]
		\begin{align}
			h(\mathbf{x}_k^i)=&\left[\begin{array}[array]{c}
				r_k^i\\
				\theta_k^i\\
				\xi_k^i
			\end{array}\right]
			=\left[\begin{array}[array]{c}
				\sqrt{\left(p^i_{x,k}- p_x^U\right)^2+\left(p^i_{y,k}- p_y^U\right)^2+\left(p^i_{z,k})- p_z^U\right)^2}\\
				\cos^{-1}\left(\frac{ p^i_{z,k}- p_z^U}{r_k^i}\right)\\
				\tan^{-1}\left({\frac{ p^i_{y,k}- p_y^U}{ p^i_{x,k}- p_x^U}}\right)
			\end{array}\right]\label{mtss}
		\end{align}
	\end{figure*}
	\subsection{Problem Formulation}
\begin{figure}[H]
		\centering
		\includegraphics[width=0.6\linewidth]{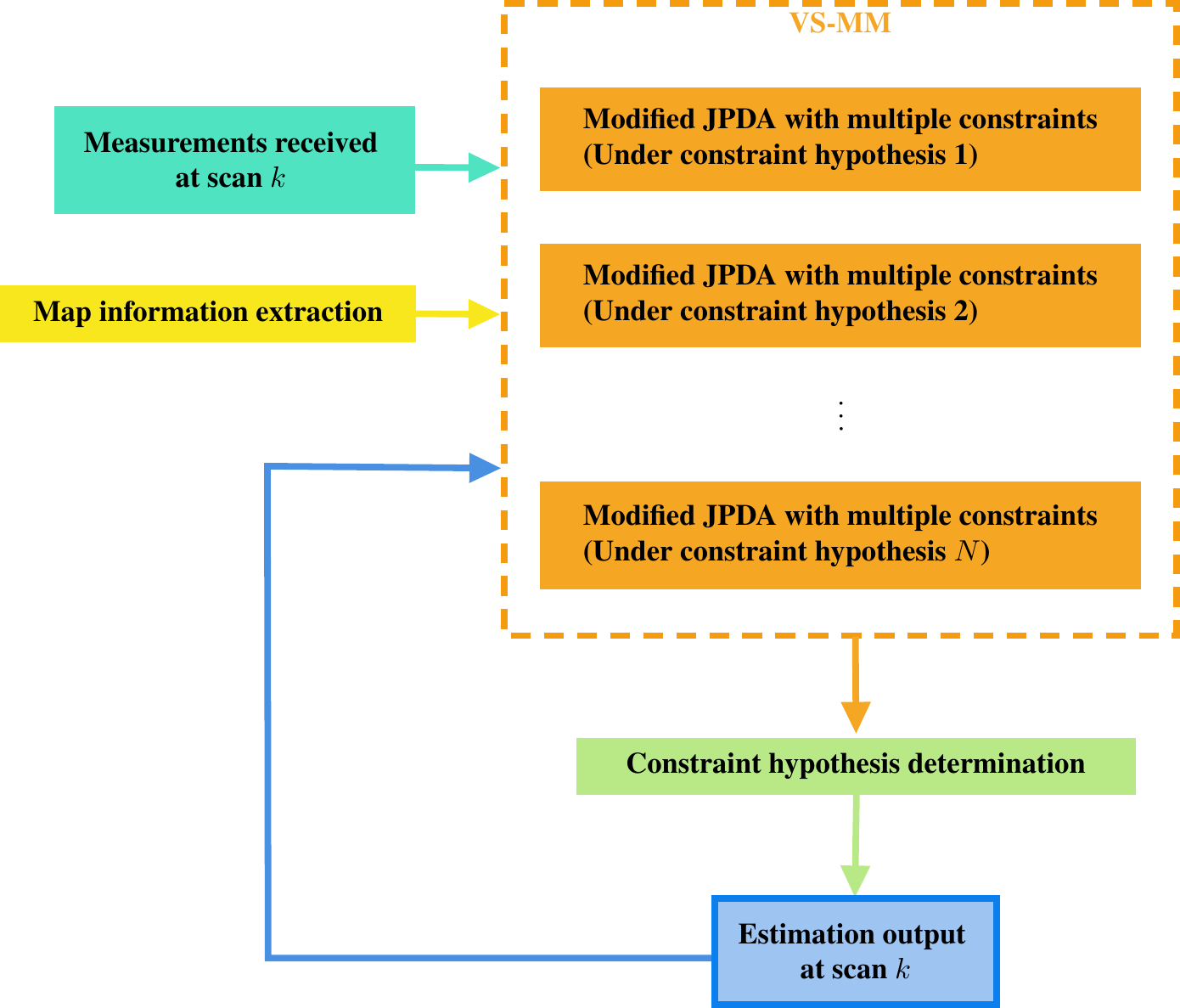}
		\caption{Logical flow of different modules.}\label{Approximated tracking logic of Modified JPDA with multiple constraints}
	\end{figure}

	The main objective of this paper is to simultaneously estimate the number of ground moving targets and the corresponding states subject to certain constraint \eqref{constraint condition} by using the UAV measurements. To solve this problem, we develop a domain-knowledge-aided airborne ground moving targets tracking algorithm that addresses three different challenges: (1) the roads from an arbitrary local map are mathematically modeled by consecutive line segments as the domain knowledge that is leveraged to cater for the on-road constraint of a ground moving vehicle; (2) conventional JPDA filter is revisited to incorporate domain knowledge as state constraints under certain constraint hypothesis, i.e., within one specific line section on the road, and a modified track update approach is also proposed to solve the problem of track loss during the monitoring process; and (3) a mathematically rigor VS-MM algorithm is suggested in consideration of multiple constraint hypotheses and find the hypothesis with the highest posterior probability for estimation output. The logical flow of different modules are presented in Fig. \ref{Approximated tracking logic of Modified JPDA with multiple constraints}.
	
	\section{Map Information Extraction}\label{sec3}
	
	This section approximates the roads from a given map by 2-D line segments that will be utilized as domain knowledge to formulate the state constraints. We first propose an algorithm to show how to approximate a given road with multiple straight-line segments and then present an application example of the proposed algorithm.
	
	%\subsection{Road Model Establishment}
	
\begin{figure}[hbt!]
		\centering
		\includegraphics[width=0.7\linewidth]{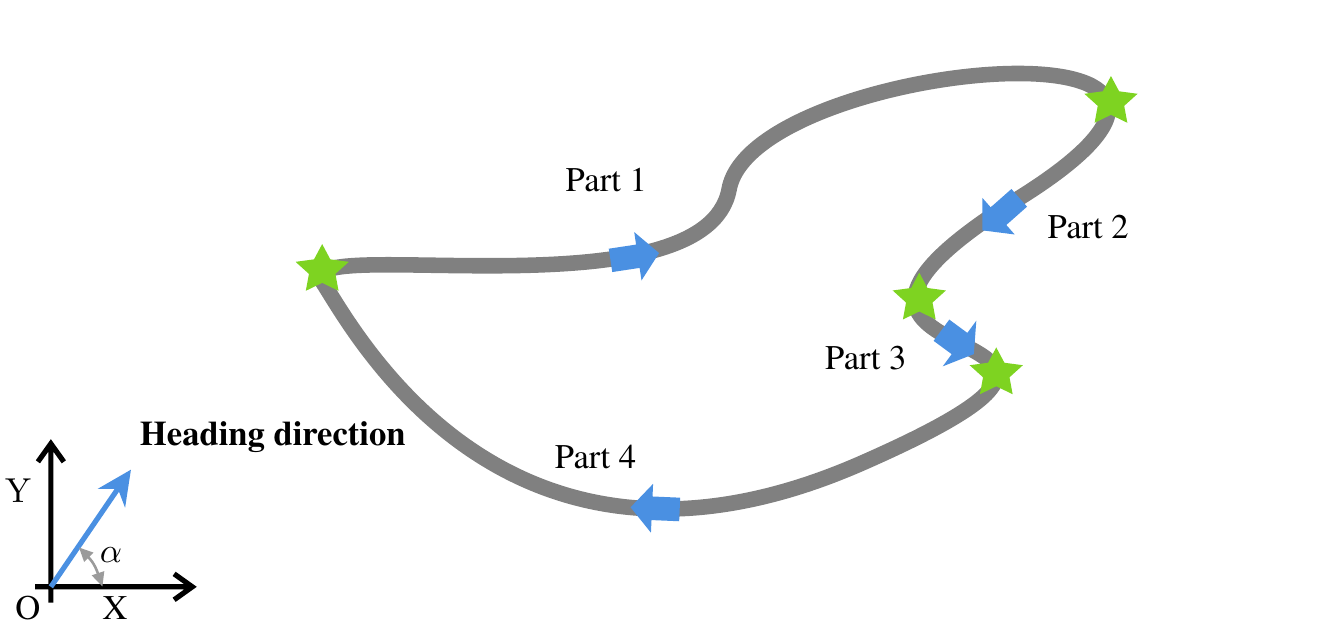}
		\caption{Road splitting depending on the heading direction.}\label{road}
	\end{figure}

	The coordinates of equally-spaced discrete geographical points which belong to the road center-line in the geographic coordinate system (GCS) are extracted and converted to the Cartesian coordinate system (CCS) through Gauss–Krüger projection \cite{projection,guass-progection}. Note that the number of discrete geographical points depends on the scale of the map and the shape of the road. The mathematical road model can then be formulated by applying polynomial fitting to the set of geographical points extracted from the map. However, it is worthy pointing out that the road shape could be extremely irregular, as shown in Fig. \ref{road}, where $\alpha\in \left(-90^{\circ},270^{\circ}\right)$ denotes the angle between the tangential direction at the discrete point and the positive direction of the $X$-axis. For the purposes of polynomial fitting, we define a new indicator, denoted by $\aleph$, to evaluate of orientation change of the road as
	\begin{align}
		\begin{cases}
			\aleph=1, &-90^{\circ}<\alpha<90^{\circ}\\
			\aleph=-1,&90^{\circ}<\alpha<270^{\circ}
		\end{cases}
	\end{align}
	
	We then split the road into several parts with each road part provides same $\aleph$, as shown in Fig. \ref{road}, where $\aleph=1$ for Parts 1, 3 and $\aleph=-1$ for Parts 2, 4. The polynomial function with order $N_p$ is leveraged to approximate one specific road part as
	\begin{align}
		f(x)=\sum\limits_{m=0}^{{N_{p}}}{{{a}_{m}}} x^m
		\label{polyfit}
	\end{align}
	where $a_m$ stands for the coefficients of the polynomial of the considered road part. Depending on the curvature of the road part, we can leverage different polynomials by tuning the value of $N_{p}$ to reach certain approximation accuracy. Obviously, increasing $N_{p}$ provides more accurate road approximation at the price of more computational burden and hence $N_{p}$ is a trade-off parameter for road approximation.
	
	\subsubsection{$N_{p}\leq1$}
	For the case where the road part is modeled as a straight-line segment, the majority of geographical information required for constraining the vehicle’s states can be obtained from two boundary points, i.e., starting point $\mathbf{s}_{s}= \left[{{x}_{s}},f({{x}_{s}})\right]^{\text{T}}$ and terminal point $\mathbf{s}_{t}=\left[{{x}_{t}},f({{x}_{t}})\right]^{\text{T}}$. For an arbitrary discrete point $\mathbf{s}=\left[x,f(x)\right]^{\text{T}}$ that locates on the straight line connecting $\mathbf{s}_{s}$ and terminal point $\mathbf{s}_{t}$, we have $\mathbf{s}\in\zeta$ with $\zeta=\left\{l\mathbf{s}_{t}+(1-l)\mathbf{s}_{s}\vert \forall l\in[0,1]\right\}$ and
	\begin{subequations}\label{pcons}
		\begin{align}
			[-\tan(\vartheta),1]\mathbf{s}=\kappa
		\end{align}
		where $\vartheta$ and $\kappa$, respectively, denote the orientation angle and the intercept of the corresponding straight line as
		\begin{align}
			&\vartheta=\arctan{\left(\frac{ f({{x}_{t}})- f({{x}_{s}})}{{{x}_{t}}-{{x}_{s}}}\right)}\\
			&\kappa=-\tan(\vartheta){{x}_{s}}+ f({{x}_{s}})
		\end{align}
	\end{subequations}
	%where $\vartheta$ denotes the orientation of the corresponding road part and will be leveraged to constrain the moving direction of the vehicle in the following section.
	
	\subsubsection{$N_{p}>1$}
	If the curvature of the road part cannot be omitted, using straight line to approximate the road might result in big approximation errors and hence it is desirable to leverage higher-order polynomials in road modeling. This, in turn, might impose nonlinear geographical constraint on target state estimation. Even though there exist some estimation algorithms nonlinear constraints reported in the literature \cite{RN10,RN7}, there are few methods that provide the capability to handle higher-order nonlinear constraints. Besides, applying nonlinear constraints for the state estimation is computationally costly, which is undesirable for small-scale UAVs in practice. Hence, approximating the curve road part by multiple straight line segments with a certain degree of precision is more practical. Given a higher-order polynomial function $f(x)$, the accumulated heading change, denoted by $\Delta\delta$, from one point $\mathbf{s}_{1}=\left[{{x}_{1}},f({{x}_{1}})\right]^{\text{T}}$ to another point $\mathbf{s}_{2}=\left[{{x}_{2}},f({{x}_{2}})\right]^{\text{T}}$ can be determined by
	\begin{align}
		\Delta\delta=\int_{{{x}_{1}}}^{x_2}{\frac{f^{\prime\prime}(x)}{1+{{[f^{ \prime }(x)]}^{2}}}dx}\label{integral}
	\end{align} 
	
	For one specific curved road part, we start from its initial point $\mathbf{s}_{s}= \left[{{x}_{s}},f({{x}_{s}})\right]^{\text{T}}$ to evaluate Eq. \eqref{integral} until $\Delta\delta>\delta_{m}$ with $\delta_{m}$ being the upper threshold of the road heading change that a straight line can approximate with certain degree of accuracy. The straight line that connects the initial point and the cutting point constitutes the first segment of the considered road part. Now, starting from the cutting point we can then find another endpoint that by evaluating Eq. \eqref{integral}. By repeating this procedure until the terminal point $\mathbf{s}_{t}=\left[{{x}_{t}},f({{x}_{t}})\right]^{\text{T}}$ of the road part is reached, we can find multiple straight line segments to approximate the road part. Once all road parts are explored, the road can be approximated by a poly-line composed of $N_{s}$ straight line segments and the $r$th segment, denoted by $\zeta_r$, is fully determined by its start point $\mathbf{s}_{r}= \left[{{x}_{r}},f({{x}_{r}})\right]^{\text{T}}$ and terminal point $\mathbf{s}_{r+1}= \left[{{x}_{r+1}},f({{x}_{r+1}})\right]^{\text{T}}$. Notice that the terminal point of the $N_s$th segment coincides with the terminal point of the road, i.e., $\mathbf{s}_{N_s+1}= \mathbf{s}_{t}$. Similar to Eq. \eqref{pcons}, the orientation angle and the intercept of the $r$th segment, denoted by $\vartheta_r$ and $\kappa_r$, are given by
	\begin{align}
		&\vartheta_r=\arctan{\frac{ f({{x}_{r+1}})- f({{x}_{r}})}{{{x}_{r+1}}-{{x}_{r}}}}\\
		&\kappa_r=-\tan(\vartheta){{x}_{r}}+ f({{x}_{r}})
	\end{align}
	
\begin{figure*}[hbt!]
		\centering
		\subcaptionbox{Road map (Kirkliston, England) extracted from Google earth: the roads between point A and point B are monitored\label{Roads map extracted from Google earth: the road between point A and point B belongs to the surveillance region}}
		{\includegraphics[width=.58\linewidth]{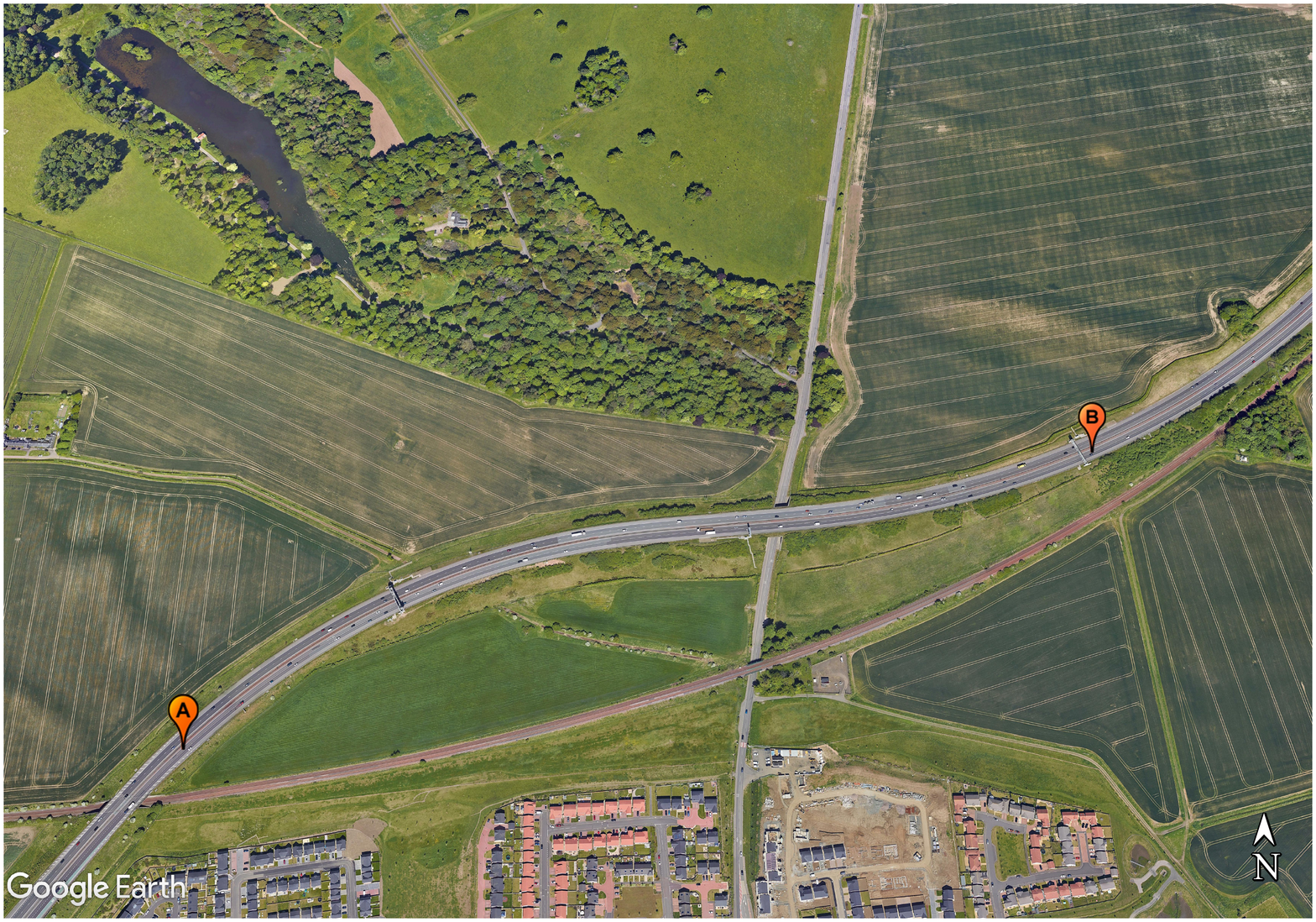}}
		\subcaptionbox{Approximating road by multiple straight-line segments \label{Approximated roads composed of multiple segments}}
		{\includegraphics[width=.7\linewidth]{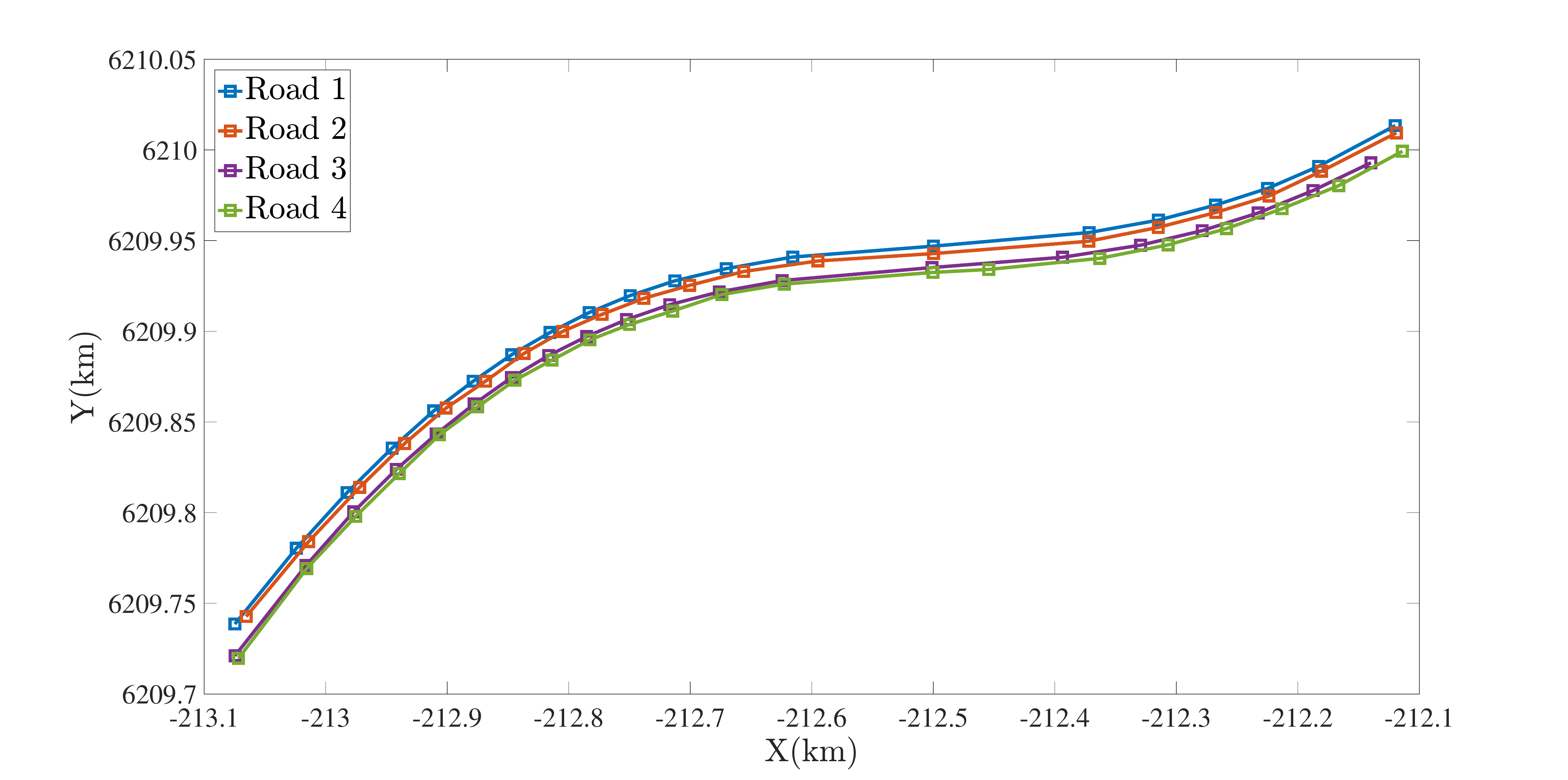}}
		\caption{Illustration example of road approximation obtained from the proposed approach.}\label{Roads approximation result}
	\end{figure*}

	An illustration of road approximation results obtained from the proposed algorithm is presented in Fig. \ref{Roads approximation result}, where the roads locate in Kirkliston and are directly taken from Google earth. From this figure, it is clear that there exist four roads in the surveillance region with two from west to east and the other two from east to west, as shown in Fig. \ref{Roads map extracted from Google earth: the road between point A and point B belongs to the surveillance region}. The tuning parameters required to implement the proposed algorithm are selected as $N_{p}=6$, $\delta_{m}=3^{\circ}$. From Fig. \ref{Approximated roads composed of multiple segments}, we can clearly observe that the intensity of nodes is proportional to the road curvature driven by the required approximation accuracy.
	
	%%%%%%%%%%%%%%%%%%%%%%%%%%%%%%%%%%%%%%%%%%%%%%%%%%%%%%%%%%%%%%%%%%%
	\section{Modified JPDA Filter with Multiple Constraints}\label{sec4}
	In this section, a modified JPDA is proposed to handle multiple constraints in state estimation, and a new tracks update logic is also developed to address the problem of track loss during target tracking. 
	
	JPDA is known as a soft data association approach that enumerates all feasible joint association hypotheses to find the target-measurement mappings. One feasible joint association hypothesis satisfies that at most one measurement originates from a target and at most one target can be assigned to a measurement. Under the assumption that all targets are mutually independent, the theoretical posterior of the $i$th target can be obtained from the Bayesian rule as \cite{RN15}
	\begin{align}
		\begin{aligned}
			p(\mathbf{x}_{k}^i\vert \chi_{k}^{i},\mathbf Z^{k})=
			&\sum_{j=0}^{M_k}p(\mathbf{x}_{k}^{i}\vert\omega^{i}_{k}=j,\chi_{k}^{i},\mathbf Z^{k})\\
			&\times p(\omega^{i}_{k}=j\vert\chi_{k}^{i},\mathbf Z^{k})\label{bayesian}
		\end{aligned}
	\end{align} 
	where $\chi_{k}^{i}$ denotes the existence of the $i$th target at scan $k$, and $\omega_k^{i}$ stands for the association event with $\omega_k^{i}=j$ meaning that the $j$th measurement is associated to the $i$th target. The term $p(\mathbf{x}_{k}^{i}\vert\omega^{i}_{k}=j,\chi_{k}^{i},\mathbf Z^{k})$ denotes the association-conditioned posterior density of the target state. This information can be extracted from a nonlinear estimation filter, e.g., EKF, in consideration of state constraints. The function $p(\omega^{i}_{k}=j\vert\chi_{k}^{i},\mathbf Z^{k})$ represents the existence-conditioned marginal association probability and can readily determined by the Bayesian rule as
	\begin{align}
		p(\omega^{i}_{k}=j\vert\chi_{k}^{i},\mathbf Z^{k})= \frac{p(\chi_{k}^{i}\vert\omega^{i}_{k}=j,\mathbf Z^{k})p(\omega^{i}_{k}=j\vert\mathbf Z^{k})}{p(\chi_{k}^{i}\vert\mathbf Z^{k})}
		\label{conde}
	\end{align}
	where the denominator is the posterior probability of target existence, which is utilized to confirm a new track and terminate an existing track. The first term in the numerator of Eq. \eqref{conde} is the association-conditioned target existence probability, which corresponds to data association status. The second term in the numerator represents conditional marginal association probability.
	
	Substituting Eq. \eqref{conde} into Eq. \eqref{bayesian} gives
	\begin{align}
		\begin{aligned}
			p(\mathbf{x}_{k}^i\vert\chi_{k}^{i},\mathbf Z^{k})=
			&\frac{1}{p(\chi_{k}^{i}\vert\mathbf Z^{k})}\sum_{j=0}^{M_k}p(\mathbf{x}_{k}^{i}\vert\omega^{i}_{k}=j,\chi_{k}^{i},\mathbf Z^{k})\\ &\times p(\chi_{k}^{i}\vert\omega^{i}_{k}=j,\mathbf Z^{k}) p(\omega^{i}_{k}=j\vert\mathbf Z^{k})
		\end{aligned}
	\end{align} 
	
	It is clear that $p(\chi_{k}^{i}\vert\mathbf Z^{k})$ can be obtained from the law of total probability with given $p(\chi_{k}^{i}\vert\omega^{i}_{k}=j,\mathbf Z^{k})$ and $p(\omega^{i}_{k}=j\vert\mathbf Z^{k})$. The proposed modified JPDA filter mainly includes three modules: calculating the marginal association probability $p(\omega^{i}_{k}=j\vert\mathbf Z^{k})$, domain-knowledge-aided state estimation $p(\mathbf{x}_{k}^{i}\vert\omega^{i}_{k}=j,\chi_{k}^{i},\mathbf Z^{k})$ and track confirmation/termination using $p(\chi_{k}^{i}\vert\mathbf Z^{k})$. Notice that enumerating all feasible joint association hypotheses is generally time-consuming and hence the ellipsoid gate technique is usually leveraged to reduce the number of feasible measurements in JPDA \cite{RN44}.

	%and a current received measurement, which is indexed by $j$ (the case $j=0$ implies no measurement received is related to the target of interest). The measurements generated by different targets are assumed detected with a independent probability $P_D$, and are selected by the "g-sigma" ellipsoid gate (please refer to for an explicit explanation), $P_G$ stands for the probability that a measurement originated from a detected target lies in the gate \cite{RN9}. The number of selected measurements by all pre-existed tracks is $M_{k}$. 

	% Thus, the original problem is transformed and can be solved by two sub-modules: \emph{Data Association} with the enumeration and calculation of feasible joint events probabilities, and \emph{Existence Probability of Tracks} with PPP birth model.  
	
	\subsection{Marginalization over All Possible Joint Association Hypotheses}\label{sec4.a}    
	%\subsubsection{Data Association}
	Each feasible joint association hypothesis in JPDA, denoted by $\Theta_k$, is essentially a measurements-to-tracks allocation set. For each pre-existed track $i=1,2,\ldots,N_{k-1}$, it is either detected or not detected at current scan $k$, and we define $\mathcal{D}$ and $\bar{\mathcal{D}}$ as the set including tracks being detected and not detected, respectively. The probability of the feasible joint association hypothesis can then be easily determined by 
	\begin{align}
		p(\Theta_k\vert\mathbf Z^{k})\propto \left(\prod_{i\in\bar{\mathcal{D}}}\bar p_{k}^{i}\right)\left(\prod_{i\in{\mathcal{D}}}p_{k}^{i}\right)
		\label{joint}
	\end{align}
	where
	\begin{align}
		\left\{
		\begin{aligned}
			&\bar p_{k}^{i}= 1-P_{D}P_{G}P(\chi_{k}^{i}\vert\mathbf Z^{k-1})\\
			&p_{k}^{i}= P_{D}P_{G}P(\chi_{k}^{i}\vert\mathbf Z^{k-1})\frac{\Lambda_{k}^{ij}}{\lambda_{e}\left(\mathbf{z}_{k}^j\right)}
		\end{aligned}
		\right.
	\end{align} 
	where $P_G$ stands for the probability that a measurement lies in the pre-designed ellipsoid gate. The notation $\Lambda_{k}^{ij}$ is the likelihood probability of the association measurement $j$ to target $i$ at scan $k$. The symbol $\lambda_{e}(\cdot)$ denotes the intensity of external source, which includes both false alarm and new born target.
	
	Similar to \cite{RN14}, we assume that new born targets are averagely distributed over the surveillance region with intensity $\lambda_{b}(\mathbf{x}_{b}^{n})$, where $\mathbf{x}_{b}^{n}$ denotes the state of the $n$th possible birth place. The intensity generated by external sources with respect to measurement $\mathbf{z}_k^j$ can then be readily obtained as 
	\begin{align}
		\lambda_{e}(\mathbf{z}_k^j)=P_{D}\sum_{n}\lambda_{b}({{\mathbf{x}}}_{b}^{n})p(\mathbf{z}_k^j\vert{{\mathbf{x}}}_{b}^{n})+\lambda_{f} \label{nbt}
	\end{align}
	
	Substituting Eq. \eqref{nbt} into Eq. \eqref{joint} and using the law of total probability, we can evaluate the marginal association probability as
	\begin{align}
		p(\omega^{i}_{k}=j\vert\mathbf Z^{k})=\sum_{\Theta_k\left(\omega_{k}^{i}=j\right)} p(\Theta_k\vert\mathbf Z^{k})\label{association}
	\end{align}
	where $\Theta_k\left(\omega_{k}^{i}=j\right)$ denotes feasible joint association hypothesis that includes $\omega_{k}^{i}=j$. 
	
	Note that the marginal association probability obtained in Eq. \eqref{association} only corresponds to confirmed or tentative tracks. Since each received measurement might originate from an external source, e.g., new born target, it is necessary to find the marginal association probability of external sources. Hence, we have
	\begin{align}\label{external}
		\begin{aligned}
			p(\omega^{i}_{k}=j\vert\mathbf{Z}^{k})= 1-\sum_{\bar i=1}^{N_{k-1}}p(\omega^{\bar i}_{k}=j\vert\mathbf{Z}^{k}),\\
			i=N_{k-1}+1,N_{k-1}+2,\ldots,N_{k-1}+M_k
		\end{aligned}
	\end{align}
	
	\subsection{State Estimation with Multiple Constraints}\label{sec4.b}
	%\subsubsection{State Estimation}
	It follows from the Bayesian theory that the state estimation is generally composed of two steps: prediction with a given dynamics model and update with received measurements, i.e.,
	\begin{align}
		p(\mathbf{x}_{k}^{i}\vert\omega^{i}_{k}=j,\chi_{k}^{i},\mathbf Z^{k})\propto \underbrace{p(\mathbf{z}_{k}(j)\vert\mathbf{x}_{k}^{i})}_{\text{Update}}\underbrace{p(\mathbf{x}_{k}^{i}\vert\chi_{k}^{i},\mathbf Z^{k-1})}_{\text{Predict}}\label{bayes}
	\end{align}
	
	Since the measurement function is nonlinear with respect to the target state, we utilize the well-known EKF for target state propagation in this paper. Suppose that the posterior of the $i$th target at time instant $t_{k-1}$ is governed by a Gaussian $\mathcal{N}\left({{\mathbf{x}}}_{k-1\vert k-1}^{i},\mathbf{P}_{k-1\vert k-1}^{i}\right)$, the association-conditioned target distribution $p(\mathbf{x}_{k}^{i}\vert \omega^{i}_{k}=j,\chi_{k}^{i},\mathbf Z^{k})$ can be obtained by following two steps as
	\begin{align}
		&\text{State Predict}: \left\{
		\begin{aligned}\label{sp}
			&{{\mathbf{x}}}_{k\vert k-1}^{i}=\mathbf{F}_{k}{{\mathbf{x}}}_{k-1\vert k-1}^{i}\\
			&\mathbf{P}_{k\vert k-1}^{i}=\mathbf{F}_{k}\mathbf{P}_{k-1\vert k-1}^{i}\mathbf{F}_{k}^{\text{T}}+\mathbf Q_{k}
		\end{aligned}
		\right.\\
		&\text{State Update}: \left\{
		\begin{aligned}
			&\mathbf{H}_{k}^{i}=\nabla_{\mathbf{x}}h(\mathbf{x})\vert_{\mathbf{x}={{\mathbf{x}}}_{k\vert k-1}^{i}}\\
			&\mathbf{S}^{i}_{k}=\mathbf{H}_{k}^{i}\mathbf{P}_{k\vert k-1}^{i}{\mathbf{H}_{k}^{i}}^{\text{T}}+\mathbf R_{k}\\
			&\mathbf{K}_{k}^{i}=\mathbf{P}_{k\vert k-1}^{i}\mathbf{H}_{k}^{i}{\mathbf{S}^{i}_{k}}^{-1}\\
			&{{\mathbf{x}}}_{k\vert k}^{ij}= {{\mathbf{x}}}_{k\vert k-1}^{i}+\mathbf{K}_{k}^{i}\left(\mathbf{z}_{k}^j-\mathbf{H}_{k}^{i}{\mathbf{x}}_{k\vert k-1}^{i}\right)\\
			&\mathbf{P}_{k\vert k}^{ij}=\left(\mathbf{I}_{6}-\mathbf{K}_{k}^{i}\mathbf{H}_{k}^{i}\right)\mathbf{P}_{k\vert k-1}^{i}
		\end{aligned}
		\right.\label{su}	
	\end{align}
	
	On substitution of Eqs. \eqref{sp} and \eqref{su} into Eq. \eqref{bayesian}, it can be easily verified that the posterior distribution of the $i$th target at time instant $t_{k}$ becomes a Gaussian mixture. However, the propagation of Gaussian mixture is computationally in tractable in practice due to the issue of curse of dimensionality. For this reason, JPDA approximates the Gaussian mixture by a single Gaussian as
	\begin{subequations}\label{GM}
		\begin{align}
			&{{\mathbf{x}}}_{k\vert k}^{i}=\sum_{j=0}^{M_{k}^{i}}\beta_{j}^{i}{{\mathbf{x}}}_{k\vert k}^{ij}\\
			&\mathbf{P}_{k\vert k}^{i}=\sum_{j=0}^{M_{k}^{i}}\beta_{j}^{i}\left[\mathbf{P}_{k\vert k}^{ij}+\left({{\mathbf{x}}}_{kv k}^{ij}-{{\mathbf{x}}}_{k\vert k}^{i}\right)\left({{\mathbf{x}}}_{k\vert k}^{ij}-{{\mathbf{x}}}_{k\vert k}^{i}\right)^{\text{T}}\right]
		\end{align}
	\end{subequations}
	where $\beta_{j}^{i}\triangleq p(\omega^{i}_{k}=j\vert \chi_{k}^{i},\mathbf Z^{k})$ and $M_{k}^{i}$ stands for the number of measurements that locate inside the ellipsoid gate of the $i$th target at scan $k$.
	%\subsubsection{Constraint Modeling and Integration}
	
	%The estimation result obtained from equation Eq. (\ref{GM}) based on EKF is not optimal, since EKF is not an optimal estimator for nonlinear system, and some approximations are made for state estimation and data association. And all these factors may add up and lead to poor tracking quality. Hence, a correction for the state estimation via incorporating additional information into estimator may help to enhance the multiple target tracking robustness.
	
	Now, we will show how to leverage the domain knowledge, i.e., road information and local traffic regulations, as state constraints to improve the performance of target estimation. For simplicity, we assume that the association between the road segment and the moving vehicle is known, i.e., the $i$th vehicle is moving on the $r$th segment at time instant $t_k$, and this association ambiguity will be addressed in Section \ref{sec5}. Under this assumption, we utilize three constraints for ground target states correction: heading direction and position constraints are utilized to ensure that the vehicle is moving on the road, and the speed limit constraint is leveraged to satisfy local traffic regulations.
	
	%According to the assumptions made in Section \ref{sec2}, there are some extra information can be considered and modeled as constraints for state estimation. Here, based on the information extracted in Section \ref{sec3}, we summarize three typical constraint models available for ground target states correction, and for illustration purposes, the match between constraints and targets are assumed known, say the state $\mathbf{x}^{i}_{k}$ corresponds to line-segment $\tau$ of road $u$ (the match confirmation will be explained thoroughly in Section \ref{sec5}).  
	
	\subsubsection{Heading Direction Constraint}
	Notice that the road is approximated by multiple straight line segments and hence the vehicle moving on the $r$th segment should align its heading direction with the heading of the segment. Obviously, this constraint can be mathematically modeled as an equality constraint as
	\begin{align}
		g_{1}({\mathbf{x}^{i}_{k}})\triangleq \left(\mathbf{x}^{i}_{k}\right)^{\text{T}}\underbrace{
			\begin{bmatrix}
				\mathbf{0}_{3\times2}\\
				\mathbf{\Delta}_{r}
		\end{bmatrix}}_{\text{Constraint matrix $\mathbf{D}^{r}_{1}$}}=\underbrace{[0,0]}_{\text{Bound $\mathbf{d}_1^r$}}
	\end{align}
	where
	\begin{align}
		\mathbf{\Delta}_r= \begin{bmatrix}
			-\tan\left(\vartheta_r\right) &1 &0\\
			0 &0 &1
		\end{bmatrix}^{\text{T}}\label{cm}
	\end{align}
	
	\subsubsection{Position Constraint}
	
	Except for the heading direction, the vehicle moving on the $r$th line segment should also satisfy the position constraint of the corresponding line segment. This constraint can also be modeled as an equality constraint as
	\begin{align}
		g_{2}({\mathbf{x}^{i}_{k}})\triangleq \left(\mathbf{x}^{i}_{k}\right)^{\text{T}}\underbrace{
			\begin{bmatrix}
				\mathbf{\Delta}_r\\
				\mathbf{0}_{3\times2}
		\end{bmatrix}}_{\text{Constraint matrix $\mathbf{D}^{r}_{2}$}}=\underbrace{\left[\kappa_r,0\right]}_{\text{Bound $\mathbf{d}_2^r$}}
	\end{align}
	where $\kappa_r$ is obtained in \eqref{pcons}.
	%where $\mathbf{\Delta}^{\tau,u}$ has been defined in Eq. (\ref{cm}), the constraint matrix is denoted by $\mathbf{D}^{\tau,u}_{2}$ (with $\mathbf{D}^{\tau,u}_{2}\in\mathbb{R}^{6\times2}$), and it is clear to testify that $\left(\mathbf{D}^{\tau,u}_{2}\right)^{\text{T}}$ has full row rank, the bound is denoted by $\mathbf{d}^{\tau,u}_{2}$ (with $\mathbf{d}^{\tau,u}_{2}\in\mathbb{R}^{2}$).
	
	\subsubsection{Speed Limit Constraint}
	
	During the traffic surveillance, the speed of the ground moving targets at a certain region is generally within a specific range due to local traffic regulations, while the speed out of the range can be considered as abnormal behaviors. The speed limit modeled is determined by an inequality constraint as
	\begin{align}
		\mathbf{v}_{\inf}\preccurlyeq g_{3}({\mathbf{x}^{i}_{k}})\triangleq \left(\mathbf{x}^{i}_{k}\right)^{\text{T}}\underbrace{
			\begin{bmatrix}
				\mathbf{0}_{3}\\
				\mathbf{I}_{3}
		\end{bmatrix}}_{\text{Constraint matrix $\mathbf{D}^{r}_{3}$}}
		\preccurlyeq\mathbf{v}_{\sup}
		\label{cm1}
	\end{align}
	where $\mathbf{v}_{\sup}\in\mathbb{R}^{3}$ denotes the supremum of the vehicle speed, and $\mathbf{v}_{\inf}\in\mathbb{R}^{3}$ denotes the infimum of the vehicle speed. Note that the vehicle speed corresponding to $Z$-direction is $0$, and thus $\mathbf{v}_{\inf}(1,3)=\mathbf{v}_{\sup}(1,3)=0$.
	
	Since directly handling the inequality constraint \eqref{cm1} in state estimation is generally difficult, we propose a projection method to relax the inequality constraint as an equality constraint: if inequality \eqref{cm1} is satisfied, the corresponding constraint is inactive; otherwise, the state is constrained to the boundary as an equality constraint. The bound is defined as $\mathbf{d}^r_{3}=\mathbf{v}_{\inf}$ if $g_{3}({\mathbf{x}^{i}_{k}}) \prec\mathbf{v}_{\inf}  $ and $\mathbf{d}^r_{3}=\mathbf{v}_{\sup}$ when $g_{3}({\mathbf{x}^{i}_{k}})\succ\mathbf{v}_{\sup}$. Combining the previous three different types of constraints, multiple sources of domain knowledge can be considered in an integrated manner by
	%\begin{align}
		%	g({\mathbf{x}_{k}^{i}})=\left(\mathbf{x}^{i}_{k}\right)^{\text{T}}\underbrace{\left[\mathbf{D}^r_{1},\mathbf{D}^r_{2},\mathbf{D}^r_{3}\right]}_{\text{Constraint matrix $\mathbf{D}_r$}}=\underbrace{\left[\mathbf{d}^r_{1},\mathbf{d}^r_{2},\mathbf{d}^r_{3}\right]}_{\text{Bound $\mathbf{d}_r$}}
		%\end{align}
	\begin{align}
		g({\mathbf{x}_{k}^{i}})\triangleq \left(\mathbf{x}^{i}_{k}\right)^{\text{T}}\mathbf{D}_r=\mathbf{d}_r
		\label{cm2}
	\end{align}
	where the constraint matrix $\mathbf{D}_r$, and the bound $\mathbf{d}_r$ are defined by considering different combinations of the constraints, as shown in Table \ref{conscmbn}.
	\begin{table*}[hbt!]\caption{Collections of multiple constraints}\label{conscmbn}
		\centering
		\begin{tabular}{@{}cccccccc@{}}
			\toprule
			\text{Case} &
			\text{1} &
			\text{2} &
			\text{3} &
			\text{4} &
			\text{5} &
			\text{6} &
			\text{7} \\ \midrule
			%		$\mathcal{C}_{k}^{i}$ &
			%		1 &
			%		2 &
			%		3 &
			%		1,2 &
			%		1,3 &
			%		2,3 &
			%		1,2,3 \\
			$\mathbf{D}_r$ &
			$\mathbf{D}^{r}_{1}$ &
			$\mathbf{D}^{r}_{2}$ &
			$\mathbf{D}^{r}_{3}$ &
			$\left[\mathbf{D}^{r}_{1},\mathbf{D}^{r}_{2}\right]$ &
			$\left[\mathbf{D}^{r}_{1},\mathbf{D}^{r}_{3}\right]$ &
			$\left[\mathbf{D}^{r}_{2},\mathbf{D}^{r}_{3}\right]$ &
			$\left[\mathbf{D}^{r}_{1},\mathbf{D}^{r}_{2},\mathbf{D}^{r}_{3}\right]$ \\
			$\mathbf{d}_r$ &
			$\mathbf{d}^{r}_{1}$ &
			$\mathbf{d}^{r}_{2}$ &
			$\mathbf{d}^{r}_{3}$ &
			$\left[\mathbf{d}^{r}_{1},\mathbf{d}^{r}_{2}\right]$ &
			$\left[\mathbf{d}^{r}_{1},\mathbf{d}^{r}_{3}\right]$ &
			$\left[\mathbf{d}^{r}_{2},\mathbf{d}^{r}_{3}\right]$ &
			$\left[\mathbf{d}^{r}_{1},\mathbf{d}^{r}_{2},\mathbf{d}^{r}_{3}\right]$ \\ \bottomrule
		\end{tabular}
	\end{table*}
	
	Inspired by the concept suggested in \cite{RN8,RN42}, we formulate an optimization problem to impose \eqref{cm2} in state estimation as follows.
	\begin{problem}
		Find the optimal state correction ${\bar{\mathbf{x}}}_{k\vert k}^{i}$ that
		\begin{equation}
			\begin{split}
				&minimize\quad\left(\bar{\mathbf{x}}^{i}_{k\vert k}-{{\mathbf{x}}}_{k\vert k}^{i}\right)^{\text{T}}{\mathbf W}^{-1}\left(\bar{\mathbf{x}}^{i}_{k\vert k}-{{\mathbf{x}}}_{k\vert k}^{i}\right)\\
				&\text{subject to\quad$\left(\bar{\mathbf{x}}^{i}_{k\vert k}\right)^{\text{T}}\mathbf{D}_r=\mathbf{d}_r$}
			\end{split}
		\end{equation}
		where $\mathbf W$ is an arbitrary symmetric positive definite weighting matrix.
	\end{problem}
	
	Notice that the objective function in Problem 1 is quadratic and the equality constraint is linear. Hence, the optimal solution can be found by solving an equivalent unconstrained optimization problem using Lagrange multiplier approach as
	\begin{align}
		\begin{aligned}
			\min_{{\bar{\mathbf{x}}}_{k\vert k}^{i},\mathbf{\lambda}}\quad &\left(\bar{\mathbf{x}}^{i}_{k\vert k}-{{\mathbf{x}}}_{k\vert k}^{i}\right)^{\text{T}}{\mathbf W}^{-1}\left(\bar{\mathbf{x}}^{i}_{k\vert k}-{{\mathbf{x}}}_{k\vert k}^{i}\right)
			\\
			&+\mathbf{\lambda}\left[\left(\bar{\mathbf{x}}^{i}_{k\vert k}\right)^{\text{T}}\mathbf{D}_r-\mathbf{d}_r\right]^{\text{T}}
		\end{aligned}
	\end{align}
	
	where $\mathbf{\lambda}$ is the Lagrange multiplier vector.
	
	Using the first-order optimality condition, we can readily derive the optimal solution as
	\begin{align}
		&{\bar{\mathbf{x}}}_{k\vert k}^{i}= \mathbf{A}_{r}{{\mathbf{x}}}_{k\vert k}^{i}+\mathbf{B}_{r}\\
		&\bar{\mathbf{P}}^{i}_{k\vert k}= \mathbf{A}_{r}\mathbf{P}_{k\vert k}^{i} {\mathbf{A}_{r}}^{\text{T}}
	\end{align}
	where the parameter matrices $\mathbf{A}_{r}$ and $\mathbf{B}_{r}$ are calculated by 
	\begin{align}	
		&\mathbf{A}_{r}= \mathbf{I}_{6}-{\mathbf W}{\mathbf{D}_r}\left({\mathbf{D}_r}^{\text{T}}{\mathbf W}{\mathbf{D}_r}\right)^{-1}{\mathbf{D}_r}^{\text{T}}\\
		&\mathbf{B}_{r}= {\mathbf W}{\mathbf{D}_r}\left({\mathbf{D}_r}^{\text{T}}{\mathbf W}{\mathbf{D}_r}\right)^{-1}{\mathbf{d}_r}^{\text{T}}
	\end{align}
	
	%Notice that system noise is supposed to be corrected meanwhile, that is, the corresponding covariance matrix $\mathbf{Q}_{k}$ should satisfy above constraint condition to generate more reasonable state prediction \cite{RN12}.
	
	%According to \emph{Remark 1.1} made in \cite{RN12}, the system noise should also be corrected and its smallest value can be obtained by,
	%\begin{align}
		%	\widetilde{Q}_{k}=\mathcal{P}^{i}_{{Q_{k}}}Q_{k}{\mathcal{P}^{i}_{{Q_{k}}}}^{\text{T}}\label{cnoise}
		%\end{align} 
	
	\subsection{Track Update Logic}\label{sec4.c}
	
	In evaluating the association-conditioned existence probability of the $i$th target, we consider three different existence hypotheses: (1) target $i$ exists and is detected; (2) target $i$ exists and is not detected; and (3) target $i$ comes from an external source. Under these three hypotheses, $p(\chi_{k}^{i}\vert \omega^{i}_{k}=j,Z^{k})$ can be determined as
	\begin{equation}
		p(\chi_{k}^{i}\vert \omega^{i}_{k}=j,\mathbf Z^{k})=
		\left\{
		\begin{split}
			&1,& \text{Hypothesis 1}\\
			&p(\chi_{k}^{i}\vert \omega^{i}_{k}=j,\mathbf Z^{k-1}),& \text{Hypothesis 2}\\
			&1-	\frac{\lambda_{f}}{\lambda_{e}\left(\mathbf{z}_{k}^j\right)},& \text{Hypothesis 3}
		\end{split}
		\right.\label{NB}
	\end{equation}
	with
	\begin{align}
		p(\chi_{k}^{i}\vert \omega^{i}_{k}=j,\mathbf Z^{k-1})=\frac{(1-P_{D}P_{G})p(\chi_{k}^{i}\vert \mathbf Z^{k-1})}{1-P_{D}P_{G}p(\chi_{k}^{i}\vert\mathbf Z^{k-1})}
	\end{align}
	where the prior probability of the existence of the $i$th pre-existed target is obtained through the Markov Chain One model \cite{RN20} with survival probability $P_s$ as
	\begin{align}
		p(\chi_{k}^{i}\vert\mathbf Z^{k-1})= P_sp(\chi_{k-1}^{i}\vert\mathbf Z^{k-1})\label{pe}
	\end{align}
	
	The posterior probability of target existence in Eq. (\ref{conde}) can be calculated by total probability law as
	\begin{align}
		p(\chi_{k}^{i}\vert\mathbf Z^{k})=\sum_{j=0}^{M_{k}^{i}}p(\chi_{k}^{i}\vert\omega^{i}_{k}=j,\mathbf Z^{k})p(\omega^{i}_{k}=j\vert\mathbf Z^{k})\label{existence}
	\end{align}
	
\begin{figure*}[hbt!]
		\centering
		\includegraphics[width=\linewidth]{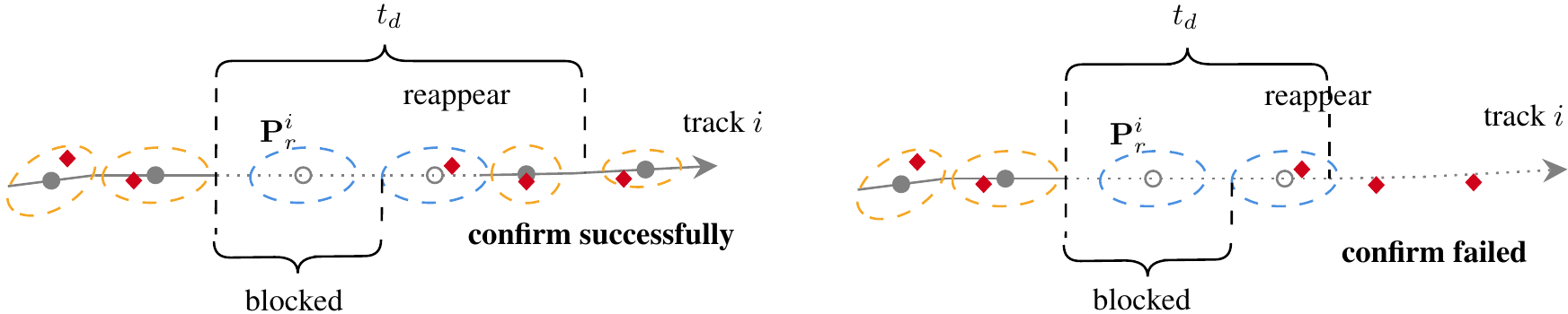}
		\caption{Illustration of reappeared target confirmation.}\label{rc}
	\end{figure*}

	Similar to \cite{RN15,RN20}, we confirm a new track if $p(\chi_{k}^{i}\vert\mathbf Z^{k})$ exceeds an upper bound $p_e$, and terminate a confirmed track if $p(\chi_{k}^{i}\vert\mathbf Z^{k})$ is smaller than a lower bound $p_t$. Although this simple logic is proved to be robust and efficient \cite{RN20}, it is challenging to recognize a temporarily disappeared target which reappears around the midpoint on the road. This can be attributed to the fact the birth place in the considered ground moving targets tracking scenario locates at the boundary of the surveillance region. To address this problem, we modify the tracking logic to realize the recognition and confirmation of reappeared tracks. The disappeared track (confirmed before) will not be removed immediately and will stay tentative with its state being predicted within a time interval threshold $t_d$. The predicted state then acts as a testing point for the selection of measurements with a predetermined covariance matrix $\mathbf{P}^{i}_{r}$. Hence, if a temporary disappeared target appears within the time interval $t_d$ and its track existence probability is bigger than $p_e$, then the track is confirmed and maintained; otherwise, it is terminated and will be removed, as shown in Fig. \ref{rc}. 
	
	%For clarity, the modified JPDA with multiple constraints is summarized in Algorithm \ref{CJPDA}.
	
	%%%%%%%%%%%%%%%%%%%%%%%%%%%%%%%%%%%%
	\section{Constraint Hypothesis Determination}\label{sec5}
	
	For the constraint JPDA algorithm proposed in the previous section, the estimation performance largely depends on determining the constraint hypothesis, i.e, which road segment the vehicle is moving on. This section proposes a VS-MM method with a hard decision logic to confirm the constraint hypothesis for existing targets and a posterior maximization approach for constraint hypothesis determination of new born targets.
	
	\subsection{Constraint Hypothesis Confirmation of Existing Targets}\label{sec5.a}
\begin{figure}
		\centering
		\includegraphics[width=0.5\linewidth]{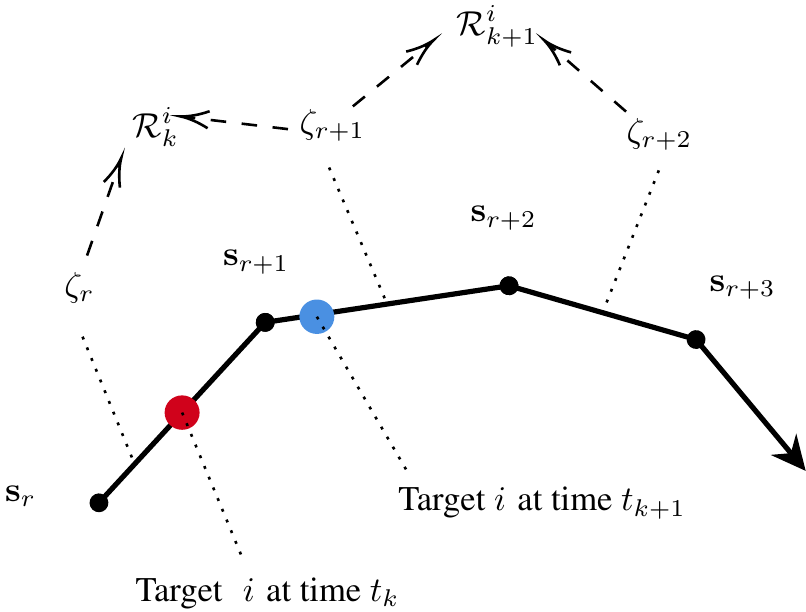}
		\caption{The allocation of target to possible road segment.}\label{mts}
	\end{figure}

	For the purpose of constraint hypothesis confirmation, we define a new variable $\eta_k^i$ and the event $\eta_k^i=\zeta_{r}$ indicates that the $i$th target is moving on road segment $\zeta_{r}$ at time instant $t_k$. Notice that only when the target of interest exists, the indicator variable $\eta_k^i$ is meaningful and hence the posterior probability of $\eta_k^i$ is defined by $p\left(\eta_k^i\vert\chi_{k}^{i},\mathbf Z^{k}\right)$. Suppose target $i$ is confirmed to be moving on the $r$th road segment at time instant $t_k$. Then, the vehicle might travel to the next adjacent segment $r+1$ at time instant $t_{k+1}$. Since the exact transformation moment between two adjacent road segments is unknown in advance and hence we define a hypothesis set $\mathcal{R}_{k}^{i}=\{\zeta_{r},\zeta_{r+1}\}$. If $p\left(\eta_{k+1}^i\vert\chi_{k+1}^{i},\mathbf Z^{k+1}\right)$ is bigger than a threshold $p_u$, the hypothesis set at time instant $t_{k+1}$ is updated as $\mathcal{R}_{k+1}^{i}=\{\zeta_{r+1},\zeta_{r+2}\}$; otherwise, we maintain $\mathcal{R}_{k+1}^{i}=\{\zeta_{r},\zeta_{r+1}\}$. The concept of allocating a target to a possible road segment is depicted in Fig. \ref{mts}. Similar to existing multiple model (MM) estimators \cite{RN16}, the prior probability of transforming from $\eta_{k-1}^i=\bar \zeta$ to $\eta_k^i=\zeta$ is determined by a transition matrix $\Pi_{\bar\zeta}^{\zeta}$ as
	\begin{align}
		\begin{aligned}
			&p\left(\eta_{k}^i=\zeta\vert\chi_{k-1}^{i},\mathbf Z^{k-1}\right)\\
			&\quad=\sum_{\bar\zeta \in \mathcal{R}_{k-1}^{i}}p(\eta_{k-1}^i=\bar\zeta\vert\chi_{k-1}^{i},\mathbf Z^{k-1})\Pi_{\bar\zeta}^{\zeta}\label{mp}
		\end{aligned}
	\end{align}
	
	Define $\Lambda_{k}^{ij}(\zeta)$ as the likelihood that the $j$th measurement is associated to the $i$th target with $\eta_{k}^i=\zeta$ at scan $k$. The total likelihood that considering two different constraint hypotheses can be readily obtained as
	\begin{align}
		\Lambda_{k}^{ij}=\sum_{\zeta\in \mathcal{R}_{k}^{i}}p\left(\eta_{k}^i=\zeta\vert\chi_{k-1}^{i},\mathbf Z^{k-1}\right)\Lambda_{k}^{ij}(\zeta)
		\label{li}
	\end{align}
	
	According to the data association presented results in Section \ref{sec4}, the posterior probability of each constraint hypothesis can be calculated as
	\begin{align}
		\begin{aligned}
			&p\left(\eta_{k}^i=\zeta\vert\chi_{k}^{i},\mathbf Z^{k}\right)\\
			&\quad\propto
			\sum_{j=0}^{M_{k}^{i}}\frac{p\left(\eta_{k}^i=\zeta\vert\chi_{k-1}^{i},\mathbf Z^{k-1}\right)\Lambda_{k}^{ij}(\zeta)}{\Lambda_{k}^{ij}}\beta_{j}^{i}
			\label{hp}
		\end{aligned}
	\end{align}
	
	If $p\left(\eta_{k}^i=\zeta\vert\chi_{k}^{i},\mathbf Z^{k}\right)$ is bigger than a given threshold $p_u$, the road transformation hypothesis is confirmed and the corresponding hypothesis set $\mathcal{R}_{k}^{i}$ is updated accordingly; otherwise, $\mathcal{R}_{k}^{i}$ remains unchanged and the probability of $\eta_{k}^i$ is propagated until condition $p\left(\eta_{k}^i=\zeta\vert\chi_{k}^{i},\mathbf Z^{k}\right)>p_u$ is triggered. Note that when the hypothesis set is updated, the posterior probability of $\eta_{k}^i$ with respect to the previous hypothesis set is invalid and shall be initialized meanwhile. 
	
	\subsection{Constraint Hypothesis Confirmation of New Born Targets}\label{sec5.b}
	A new born target is then initialized as a mixture of Gaussian distributions $\mathcal N \left({\mathbf x}_{b}^{n},\mathbf{P}_{b}^{n}\right)$ as
	\begin{subequations}\label{new}
		\begin{align}
			&{{\mathbf{x}}}_{k\vert k}^{i}=\sum_{n}\beta_{j}^{n}{{\mathbf{x}}}_{b}^{n}\\
			&\mathbf{P}_{k\vert k}^{i}=\sum_{n}\beta_{j}^{n}\left[\mathbf{P}_{b}^{n}+\left({{\mathbf{x}}}_{b}^{n}-{{\mathbf{x}}}_{k\vert k}^{i}\right)\left({{\mathbf{x}}}_{b}^{n}-{{\mathbf{x}}}_{k\vert k}^{i}\right)^{\text{T}}\right]
		\end{align}
	\end{subequations}
	where $\beta_{j}^{n}$ denotes the association probability that measurement $j$ is assigned to the $n$th birth place and is given by
	\begin{align}
		\beta_{j}^{n}\propto P_D\lambda_{b}({{\mathbf{x}}}_{b}^{n})p(\mathbf{z}_{k}^{j}\vert{{\mathbf{x}}}_{b}^{n})
	\end{align}
	
	Since the targets are moving on a road, the state of each birth place should be constrained on a specific road segment. Among all possible birth places on the road, we utilize maximum posterior criterion to find the most probable for birth target initialization as
	\begin{align}
		\arg\max_{n}\beta_{j}^{n}
	\end{align}
	
	\subsection{Constraint Hypothesis Conditioned Estimation}\label{sec5.c}
	
	From Fig. \ref{mts}, it is clear that each target corresponds to two constraint hypotheses at each scan, and the two corresponding constraint JPDA filters, which are running in parallel, can be considered as a MM estimator.  Since hypothesis set $\mathcal{R}_{k}^{i}$ varies with time, the proposed multiple model estimator is termed as a VS-MM method. Under a given constraint hypothesis, the posterior of $i$th target in Eq. \eqref{bayesian} can be reformulated as
	\begin{align}
		\begin{aligned}
			p(\mathbf{x}_{k}^i\vert\chi_{k}^{i},\eta_{k}^i,\mathbf Z^{k})= &\sum_{j=0}^{M_k^i}p(\omega^{i}_{k}=j\vert\chi_{k}^{i},\eta_{k}^i,\mathbf Z^{k})\\
			&\times p(\mathbf{x}_{k}^{i}\vert\omega^{i}_{k}=j,\chi_{k}^{i},\eta_{k}^i,\mathbf Z^{k})
			\label{bayesian2}
		\end{aligned}
	\end{align} 
	where the constraint hypothesis conditioned marginal probability $p(\omega^{i}_{k}=j\vert\chi_{k}^{i},\eta_{k}^i,\mathbf Z^{k})$ is determined by the Bayesian rule as 
	\begin{align}\label{marg}
		\begin{aligned}
			&p(\omega^{i}_{k}=j\vert\chi_{k}^{i},\eta_{k}^i,\mathbf Z^{k})\\ &=p(\omega^{i}_{k}=j\vert\chi_{k}^{i},\mathbf Z^{k})\frac{p\left(\eta_{k}^i\vert\chi_{k-1}^{i},\mathbf Z^{k-1}\right)\Lambda_{k}^{ij}(\eta_{k}^i)}{p\left(\eta_{k}^i\vert\chi_{k}^{i},\mathbf Z^{k}\right)\Lambda_{k}^{ij}}
		\end{aligned}
	\end{align}
	
	In Eq. \eqref{bayesian2}, we can extract the mean ${{\mathbf{x}}}_{k\vert k}^{ij}({\eta_{k}^{i}})$ and its corresponding covariance ${{\mathbf{P}}}_{k\vert k}^{ij}({\eta_{k}^{i}})$ of the hypothesis and association conditioned probability density $p(\mathbf{x}_{k}^{i}\vert \omega^{i}_{k}=j,\chi_{k}^{i},\eta_{k}^i,\mathbf Z^{k})$ using EKF, as shown in Eqs. \eqref{sp} and \eqref{su}. Notice that Eq. \eqref{bayesian2} is Gaussian mixture and hence we also use a single Gaussian to approximate the mixture for the purpose of reducing computational complexity as
	approximates the Gaussian mixture by a single Gaussian in Eq. \eqref{GM}
	\begin{figure*}[htp!]
		\begin{subequations}\label{GM}
			\begin{align}
				&{{\mathbf{x}}}_{k\vert k}^{i}({\eta_{k}^{i}})=\sum_{j=0}^{M_{k}^{i}}\beta_{j}^{i}({\eta_{k}^{i}}){{\mathbf{x}}}_{k\vert k}^{ij}({\eta_{k}^{i}})\\
				&\mathbf{P}_{k\vert k}^{i}({\eta_{k}^{i}})=\sum_{j=0}^{M_{k}^{i}}\beta_{j}^{i}({\eta_{k}^{i}}) \left[\mathbf{P}_{k\vert k}^{ij}({\eta_{k}^{i}})+\left({{\mathbf{x}}}_{k\vert k}^{ij}({\eta_{k}^{i}})-{{\mathbf{x}}}_{k\vert k}^{i}({\eta_{k}^{i}})\right)\left({{\mathbf{x}}}_{k\vert k}^{ij}({\eta_{k}^{i}})-{{\mathbf{x}}}_{k\vert k}^{i}({\eta_{k}^{i}})\right)^{\text{T}}\right]
			\end{align}
		\end{subequations}
		with $\beta_{j}^{i}({\eta_{k}^{i}})\triangleq p(\omega^{i}_{k}=j\vert \chi_{k}^{i},\eta_{k}^i,\mathbf Z^{k})$.
	\end{figure*}
	
	Notice that the proposed VS-MM estimator runs two JPDA in parallel and hence stores two different estimates for each target at every time instant. For the output to the users, it follows from Fig. \ref{mts} that it would be wise to choose the first element of $\mathcal{R}_{k}^{i}$, i.e., $\mathcal{R}_{k}^{i}(1)$, as the trusted mode since $\mathcal{R}_{k}^{i}(1)$ stands for the most probable road segment that the vehicle is currently moving on. By summarizing the results presented in previous sections, the proposed ground moving target tracking approach is presented in Algorithm \ref{MM}. 
	
	\begin{breakablealgorithm}
		\caption{Domain-Knowledge-Aided Airborne Ground Targets Tracking}\label{MM}
		\centering
		\begin{algorithmic}[1]
			\Require{Currently received measurements $\mathbf{Z}_{k}$, prior estimation results $\left\{\bar{{\mathbf{x}}}_{k-1\vert k-1}^{i},\bar{\mathbf{P}}_{k-1\vert k-1}^{i}\right\}$ with $i\in\left\{1,\ldots,N_{k-1}\right\}$, domain knowledge}
			\Ensure{State estimation after correction $\left\{\bar{{\mathbf{x}}}_{k\vert k}^{i},\bar{\mathbf{P}}_{k\vert k}^{i}\right\}$ with $i\in\left\{1,\ldots,N_{k}\right\}$}
			\If{$N_{k-1}\neq0$}
			\Statex \Comment Estimation for pre-existed targets
			\For{$i=1:N_{k-1}$}
			\State Get prior density $p(\chi_{k}^{i}\vert \mathbf Z^{k-1})\gets$ Eq. (\ref{pe})
			\State $p\left(\eta_{k}^i\vert \chi_{k-1}^{i},\mathbf Z^{k-1}\right)\gets$  Eq. (\ref{mp})
			\For{ each $\eta_{k}^i\in\mathcal{R}_{k-1}^{i}$}
			%						\tcc{state estimation}
			\State Apply state prediction process in Sec. \ref{sec4.b}
			\State Get $M_{k}^{i}$ through measurement gating and calculate the likelihood $\Lambda_{k}^{ij}(\eta_{k}^i)$ for each $j$
			\State Update target's state in Sec. \ref{sec4.b} to obtain $\left\{{{\mathbf{x}}}_{k\vert k}^{ij}(\eta_{k}^{i}),\mathbf{P}_{k\vert k}^{ij}(\eta_{k}^{i})\right\}$
			\EndFor
			\State $\Lambda_{k}^{ij}\gets$ Eq. (\ref{li})
			\EndFor						
			\Statex \Comment Data association
			\State Enumerate feasible joint events and calculate corresponding probability with $p(\Theta_{k}\vert Z^{k})\gets$ Eq. (\ref{joint})
			\For{$i=1:N_{k-1}$}
			\State Calculate marginal association probability $\beta_{j}^{i}$ for each $j\in\left\{1,\ldots,M_{k}^{i}\right\}$ in Sec. \ref{sec4.a} and update track existence status simultaneously in Sec. \ref{sec4.c}
			\State Calculate posterior density of constraint hypothesis $p\left(\eta_{k}^i\vert \chi_{k}^{i},\mathbf Z^{k}\right)$ in Sec. \ref{sec5.a}
			\State Get $\beta_{j}^{i}(\eta_{k}^{i})$ for each $j\in\left\{1,\ldots,{M_{k}^{i}}\right\}$ and obtain $\left\{{{\mathbf{x}}}_{k\vert k}^{i}(\eta_{k}^{i}),\mathbf{P}_{k\vert k}^{i}(\eta_{k}^{i})\right\}$ through Gaussian mixture approximation in Sec. \ref{sec5.c} $\text{ for each $\eta_{k}^{i}$}$
			\EndFor
			\EndIf
			\Statex \Comment{Track new born targets and update track status}
			\State Calculate new born track probability $p(\chi_{k}^{i}\vert Z^{k})$ in Sec. \ref{sec4.c}, for each received measurement $j=1,\ldots,M_{k}$
			\State Update track existence status and get $N_{k}$ by applying track update logic in Sec. \ref{sec4.c}
			\State Initialize the state of new born tracks in Sec. \ref{sec5.b}
			\For{$i=1:N_{k}$}
			\Statex \Comment{Constraint hypothesis determination}
			\State Update constraint hypothesis set $\mathcal{R}_{k}^{i}$ in Sec. \ref{sec5.a}
			\State Get $\left\{\bar{{\mathbf{x}}}_{k\vert k}^{i},\bar{\mathbf{P}}_{k\vert k}^{i}\right\}$ with respect to each hypothesis in $\mathcal{R}_{k}^{i}$ by correcting target state in Sec. \ref{sec4.b}
			\State Select $\mathcal{R}_{k}^{i}(1)$ as the trusted constraint hypothesis and the corresponding estimate as output to users
			\EndFor\\
			\Return{$\left\{\bar{{\mathbf{x}}}_{k\vert k}^{i},\bar{\mathbf{P}}_{k\vert k}^{i}\right\}$ for each $i\in\left\{1,\ldots,N_{k}\right\}$}
		\end{algorithmic}
	\end{breakablealgorithm}
	
	%%%%%%%%%%%%%%%%%%%%%%%%%%%%%%%%%%%%
	\section{Numerical Simulations}\label{sec6}
	
	In this section, the performance of proposed multiple target tracking algorithm is investigated through Monte Carlo simulations, and the impact of some important tuning parameters on the tracking performance is evaluated via comparison. 
	
	\subsection{Simulation Setup}
	
\begin{figure*}[hbt!]
		\centering
		{\includegraphics[width=.9\linewidth]{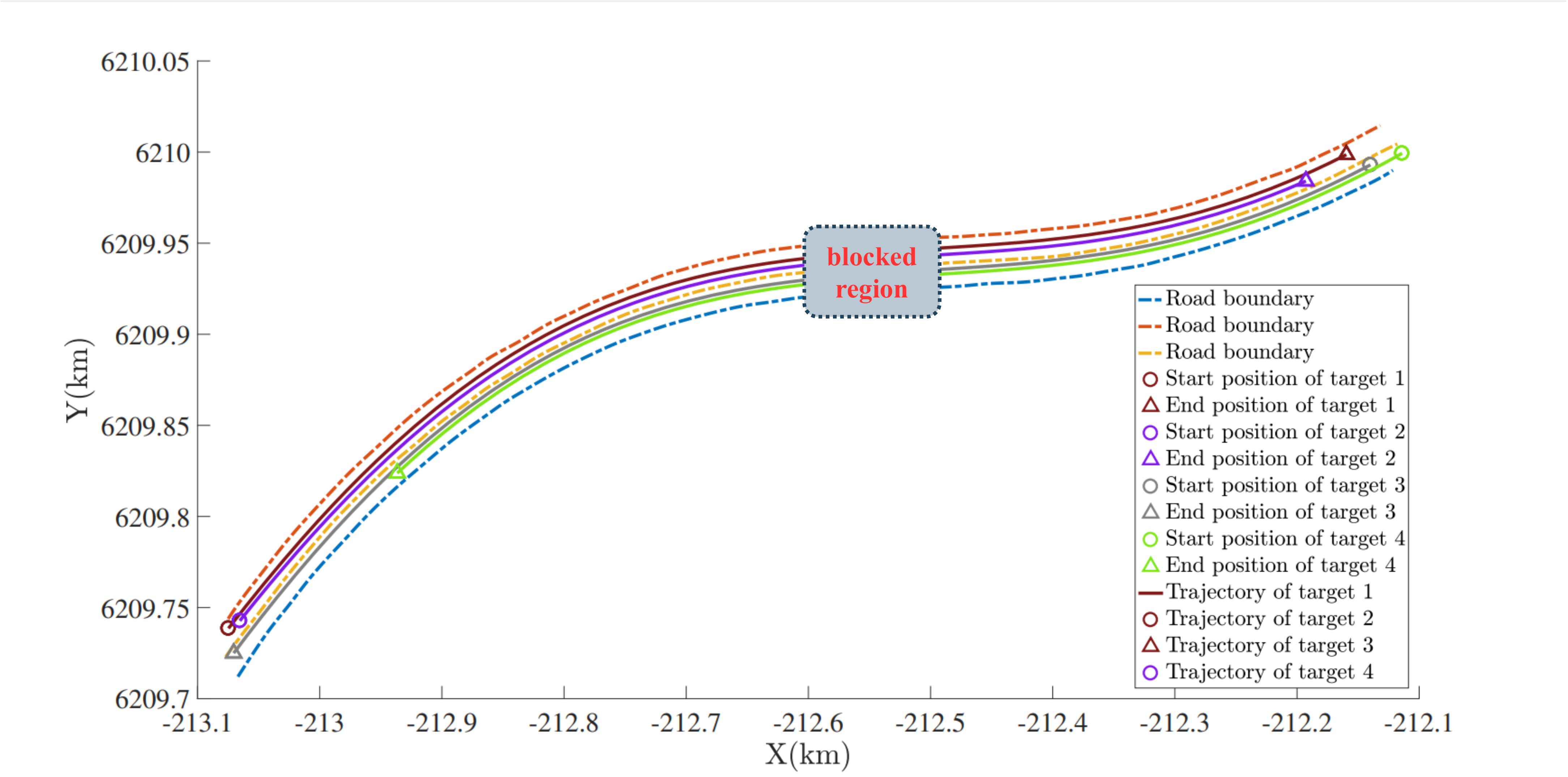}}
		\caption{Aerial surveillance scenario considered in the simulations.}\label{truetrack}
	\end{figure*}

	We establish a multiple target tracking scenario, involving 4 vehicles moving on 4 curve roads and monitored by a hovering drone. The local topography information is extracted in Section \ref{sec3}. The four vehicles enter the surveillance region from the edge of the roads at different time instants and disappear with different terminal time, i.e., 1s$\sim$57s, 25s$\sim$80s,10s$\sim$66s and 30s$\sim$80s. The velocity of each vehicle changes dynamically and remains in a certain range which is assumed to be known. The scenario depicted above is shown in Fig. \ref{truetrack}. The parameters required in the simulations, unless otherwise specified, are shown in Table \ref{pv}. Since there is a four-lane road with 2 directions from the local map, we choose four possible birth places located at the boundaries of the roads as $\mathbf{x}_{b}^{1}=[-213074.6991,6209735.658,0,0,0,0]^{\text{T}}$, $\mathbf{x}_{b}^{2}=[-213065.3738,6209739.889,0,0,0,0]^{\text{T}}$, $\mathbf{x}_{b}^{3}=[-212139.8879, 6209992.308,0,0,0,0]^{\text{T}}$, and $\mathbf{x}_{b}^{4}=[-212113.9793, 6209997.771,0,0,0,0]^{\text{T}}$. The corresponding covariance matrices are $\mathbf{P}_{b}^{1}=\mathbf{P}_{b}^{2}=\mathbf{P}_{b}^{3}=\mathbf{P}_{b}^{4}=\text{diag}([50,50,50,141,141,10])$.
	
	% \usepackage{booktabs}
	\begin{table*}[hbt!]
		\centering
		\caption{Parameters required in the simulations}
		\label{pv}
		\begin{tabular}{@{}ll@{}}
			\toprule\toprule
			\textbf{Parameter}                           & \textbf{Value}         \\ \midrule
			Simulation length (s)                         & 80                     \\
			Sampling time $t_{k\vert k-1}$ (s)                 & 1                      \\
			Velocity range (m/s)                   & [11,23]                 \\
			System noise $\mathbf Q_{k}$ & diag([20,20,10,20,20,4]) \\
			Sensor noise $\mathbf R_{k}$                & diag([1,1,1])            \\
			Average number of clutter                    & 20                     \\
			Detection probability $P_D$                         & 0.95                   \\
			Gating probability   $P_G$                         & 0.99                   \\
			Location of UAV (m)                          &$[-213224.699,6209585.65,150]^{\text{T}}$    \\ 
			Track confirmation threshold $p_{e}$         & 0.8                    \\
			Track termination threshold $p_{t}$                 & 0.2                    \\ 
			Constraint hypothesis confirmation threshold $p_{u}$  & 0.65                    \\
			Constraint type                              &Case 4 in Table \ref{conscmbn} \\
			Maximum heading change $\delta_{m}$ (deg)     & 3                  \\
			Reconfirmation covariance matrix $\mathbf{P}^{i}_{r}$ &diag([120,120,50,500,500,5])           \\
			Reconfirmation time threshold $t_d$ (s)         & 20               \\
			Weighting matrix $\mathbf W$                           & $\mathbf{I}_{6}$               \\
			Number of Monte Carlo runs                      & 300              \\
			\bottomrule\bottomrule
		\end{tabular}
	\end{table*}
	
	%The numerical simulation mainly consists of four components, which test the effectiveness of proposed algorithm from various angles. The first work shows the impact of roads segmentation introduced in Section \ref{sec3} to target tracking quality. The second part aims to testify the effectiveness of hypothesis determination and update module mentioned in Section \ref{sec5}, and shows the simulation results of target tracking via comparison between different value of hypothesis update confirm threshold. Then, the automatism of proposed algorithm for multiple target tracking is tested by including obstruction into surveillance region, and the final work evaluates the tracking performance of those multiple target tracking filters with different constraint collections, and illustrates the point that integration of multiple constraints helps to enhance the tracking performance through comparison.    
	%\subsection{Performance Metric}
	Notice that the number of targets at the surveillance region changes dynamically. This causes ambiguity using conventional metrics such as Root Mean Squared Error (RMSE) or Mean Squared Error (MSE) in performance evaluation. An overall performance metric named Optimal Subpattern Assignment (OSPA) metric, proposed in \cite{RN5}, is utilized to evaluate both the localization and cardinality estimation performance. For two vector sets $\mathbf X\sim\{\mathbf x_{1},\ldots,\mathbf x_{m}\}$, $\mathbf Y\sim\{\mathbf y_{1},\ldots,\mathbf y_{n}\}$, the localization and cardinality errors are defined in Eq. \eqref{merr}.
	\begin{figure*}
		\begin{subequations}\label{merr}
			\begin{align}
				%	\begin{array}{l}
					e_{p, l o c}^{(c)}(\mathbf X, \mathbf Y):=\begin{cases}
						\left\{\frac{1}{n} \min _{\pi \in \Pi_{n}} \sum_{i=1}^{m} d^{(c)}\left(\mathbf x_{i}, \mathbf y_{\pi(i)}\right)^{p}\right\}^{1 / p},&\text{if $m\leq n$}\\
						e_{p, l o c}^{(c)}(\mathbf Y, \mathbf X),&\text{otherwise}
					\end{cases}
				\end{align}
				\begin{align}
					e_{p, c a r d}^{(c)}(\mathbf X, \mathbf Y):=\begin{cases}
						\left[\frac{c^{p}(n-m)}{n}\right]^{1 / p},&\text{if $m\leq n$}\\
						e_{p, c a r d}^{(c)}(\mathbf Y, \mathbf X),&\text{otherwise}
					\end{cases}
					%\end{array}
			\end{align}
		\end{subequations}
	\end{figure*}
	
	where $d^{(c)}\left(\mathbf x_{i}, \mathbf y_{\pi(i)}\right)^{p}=\min\left(c, d\left(\mathbf x_{i}, \mathbf y_{\pi(i)}\right)\right)$ denotes the cut-off distance between two vectors. The order parameter $p$ and cut-off parameter $c$ are set as $c=25$ and $p=1$ in following simulations. $\Pi_{n}$ is the set of permutations on $\{1,2,\ldots,n\}$ for an arbitrary positive integer $n$.
	
	The performance metric, known as the OSPA distance, takes both localization error and the cardinality error into account is defined in Eq. \eqref{ospa}.
	\begin{figure*}
		\begin{align}\label{ospa}
			d^{(c)}_{p}(\mathbf X,\mathbf Y):=\begin{cases}
				\left\{\left(e_{p, l o c}^{(c)}(\mathbf X, \mathbf Y)\right)^{p}+\left(e_{p, card}^{(c)}(\mathbf X, \mathbf Y)\right)^{p}\right\}^{1 / p},&\text{if $m\leq n$}\\
				d^{(c)}_{p}(\mathbf Y,\mathbf X),&\text{otherwise}
			\end{cases}
		\end{align}
	\end{figure*}
	
	%
	%OSPA defines two critical parameters: the order parameter $p$ and cut-off parameter $c$, whose functions are thoroughly explained in , and in following simulations, these two parameters stay constant and are assigned as: $c=25$ and $p=1$. If $p<\infty$,
	
	\subsection{Performance with Different Road Segmentations}
	%\subsubsection{The Impact of Road Segmentation to Tracking Results}
	To extract domain knowdge constraint information, the roads are approximated by a certain number of linear segments in Section \ref{sec3}. The number of line-segments, determined by the maximum heading change $\delta_{m}$, is essential to the tracking quality as it directly impacts the precision of constraint modeling. To investigate the effect of road segmentation module on the tracking performance, we evaluate the proposed algorithm with various $\delta_{m}=1^{\circ}, 3^{\circ}, 5^{\circ}, 10^{\circ}$.
	
\begin{figure*}[hbt!]
		\centering
		\subcaptionbox{Simulation results of mean OSPA distance\label{simulation1-d}}
		{\includegraphics[width=.45\linewidth]{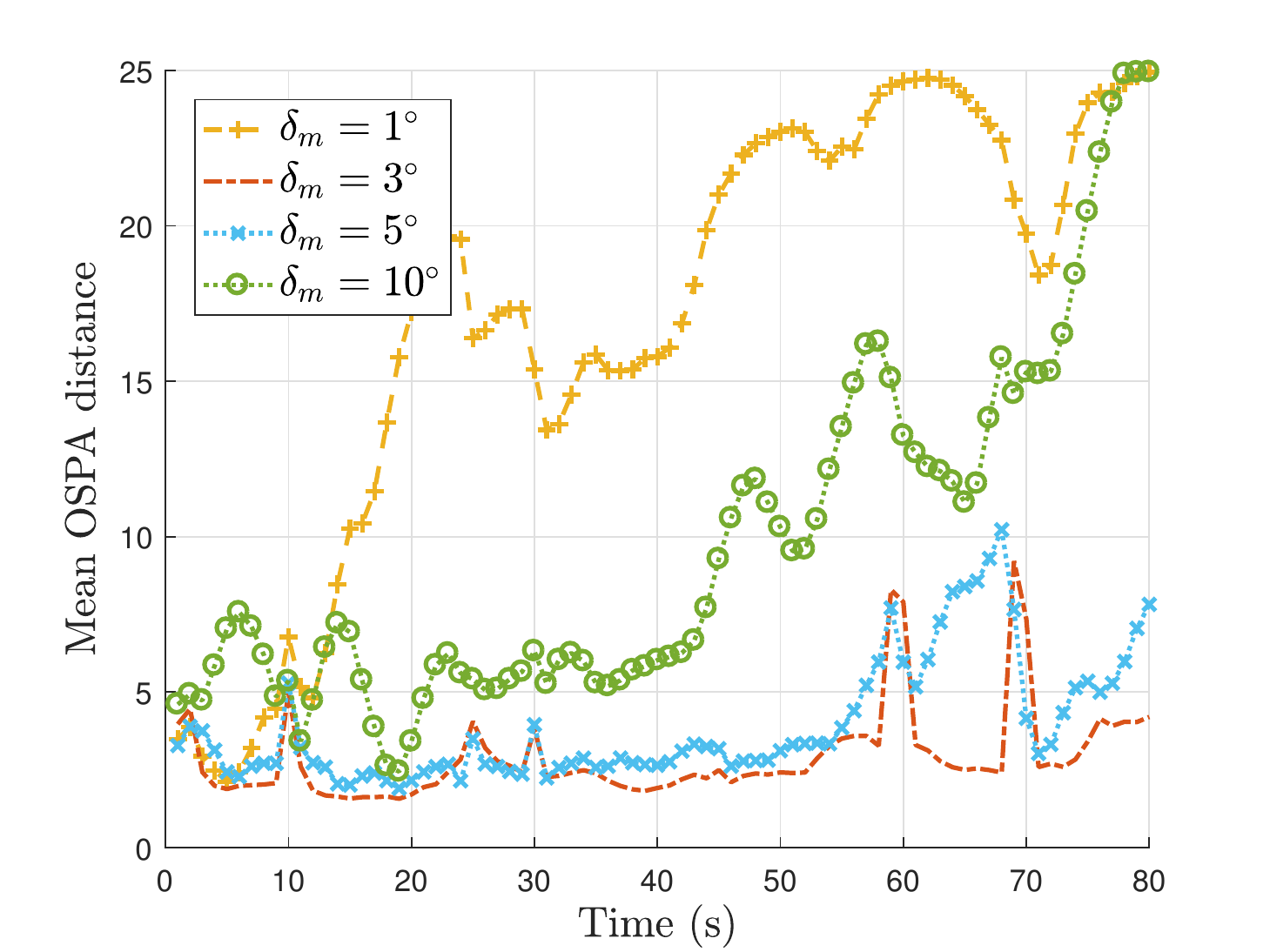}}
		\subcaptionbox{Simulation results of mean cardinality\label{simulation1-c}}
		{\includegraphics[width=.45\linewidth]{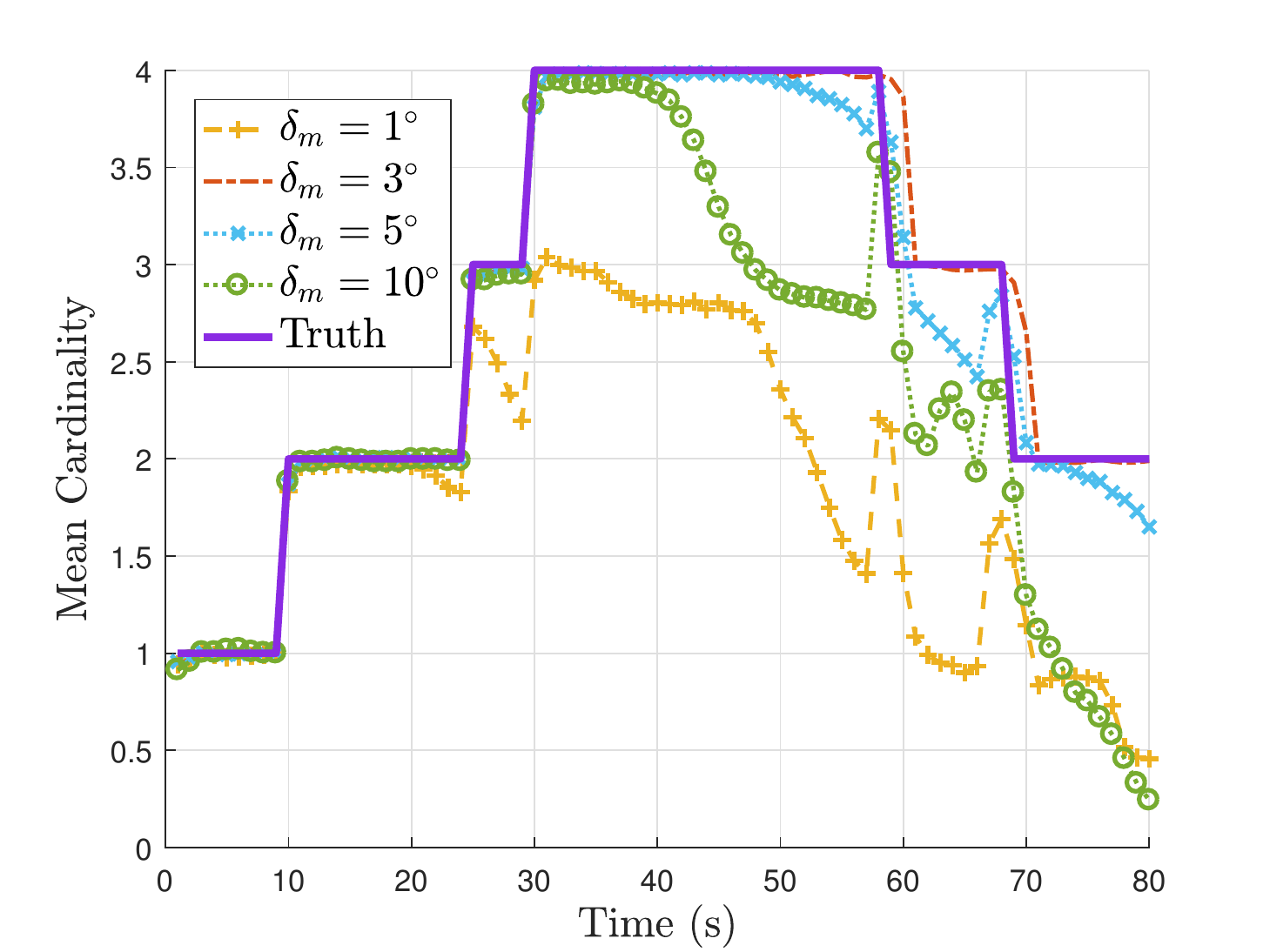}}
		\caption{Simulation results of OSPA with various $\delta_{m}$.}\label{simulation1}
	\end{figure*}

	The simulation results of mean OSPA distance and mean cardinality obtained from the proposed algorithm are shown in Fig. \ref{simulation1}, and statistics of the estimation errors are summarized in Table \ref{simulation1-tb}. The results clearly reveal that the proposed algorithm with ${{\delta }_{m}}=3^{\circ}$ generates better estimation performance than the other three parameter settings. With bigger value of $\delta_{m}$, e.g., $\delta_{m}=10^{\circ}$, the road segmentation is sparse and straight line approximation errors cannot be omitted, hence providing poor tracking quality. On the contrary, when the value of ${\delta }_{m}$ is too small, e.g., ${{\delta }_{m}}=1^{\circ}$ shown in Fig. \ref{simulation1}, the tracking performance degrades dramatically. This is mainly attributed to the failure of constraint hypothesis determination, i.e., the incorrect allocations of segment to targets. The quality can be improved by adjusting the constraint hypothesis confirmation threshold $p_u$.
	\begin{table*}[hbt!]
		\centering
		\caption{Statistics of estimation errors with various $\delta_{m}$}\label{simulation1-tb}
		\begin{tabular}{@{}ccccccc@{}}
			\toprule
			& \multicolumn{2}{c}{\text{OSPA distance}}      & \multicolumn{2}{c}{\text{Localization error}} & \multicolumn{2}{c}{\text{Cardinality error}}  \\ \cmidrule(l){2-7} 
			\multirow{-2}{*}{$\delta_{m}$} & Mean    & Std    & Mean   & Std    & Mean   & Std    \\ \midrule
			$1^{\circ}$                                & 17.0560 &7.0093& 9.1759 &4.1363 & 7.8801 &5.9172  \\
			$3^{\circ}$                                & \textbf{2.8962} &\textbf{1.4498}& \textbf{2.2350} &\textbf{0.5758}& \textbf{0.6612} &1.3300 \\
			$5^{\circ}$                                & 3.9127 &1.9625  & 2.7975 &1.0442& 1.1152 &\textbf{1.2083}  \\
			$10^{\circ}$                               & 9.7210 &5.6316   &5.1250 &1.8880&4.5959  &5.4373 \\ \bottomrule
		\end{tabular}
	\end{table*}
	%\midrule
	
	\subsection{Performance with Different Constraint Hypothesis Determination Thresholds}
	Notice that the constraint hypothesis set utilized to apply the correct domain knowledge to improve the estimation accuracy, and is updated if condition $p\left(\eta_{k}^i=\zeta\vert \chi_{k}^{i},\mathbf Z^{k}\right)>p_u$ is satisfied. Thus, we investigate the effect of $p_u$ with five different values, i.e., $p_u=0.55$, $p_u=0.61$, $p_u=0.65$, $p_u=0.71$ and $p_u=0.76$, on target tracking quality. 
	
\begin{figure*}[hbt!]
		\centering
		\subcaptionbox{Simulation results of mean OSPA distance \label{simulation2-d}}
		{\includegraphics[width=.45\linewidth]{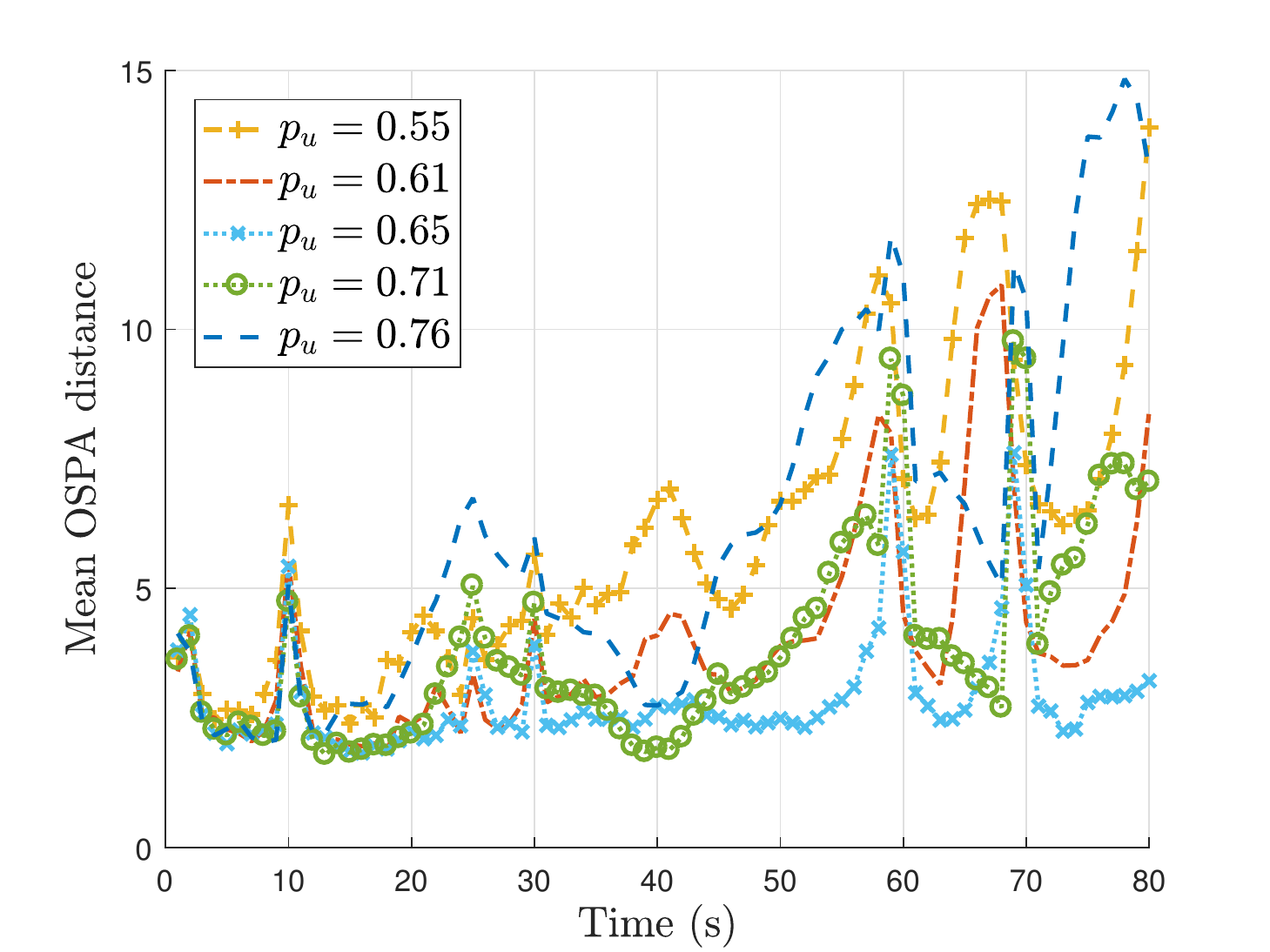}}
		\subcaptionbox{Simulation results of mean cardinality\label{simulation2-c}}
		{\includegraphics[width=.45\linewidth]{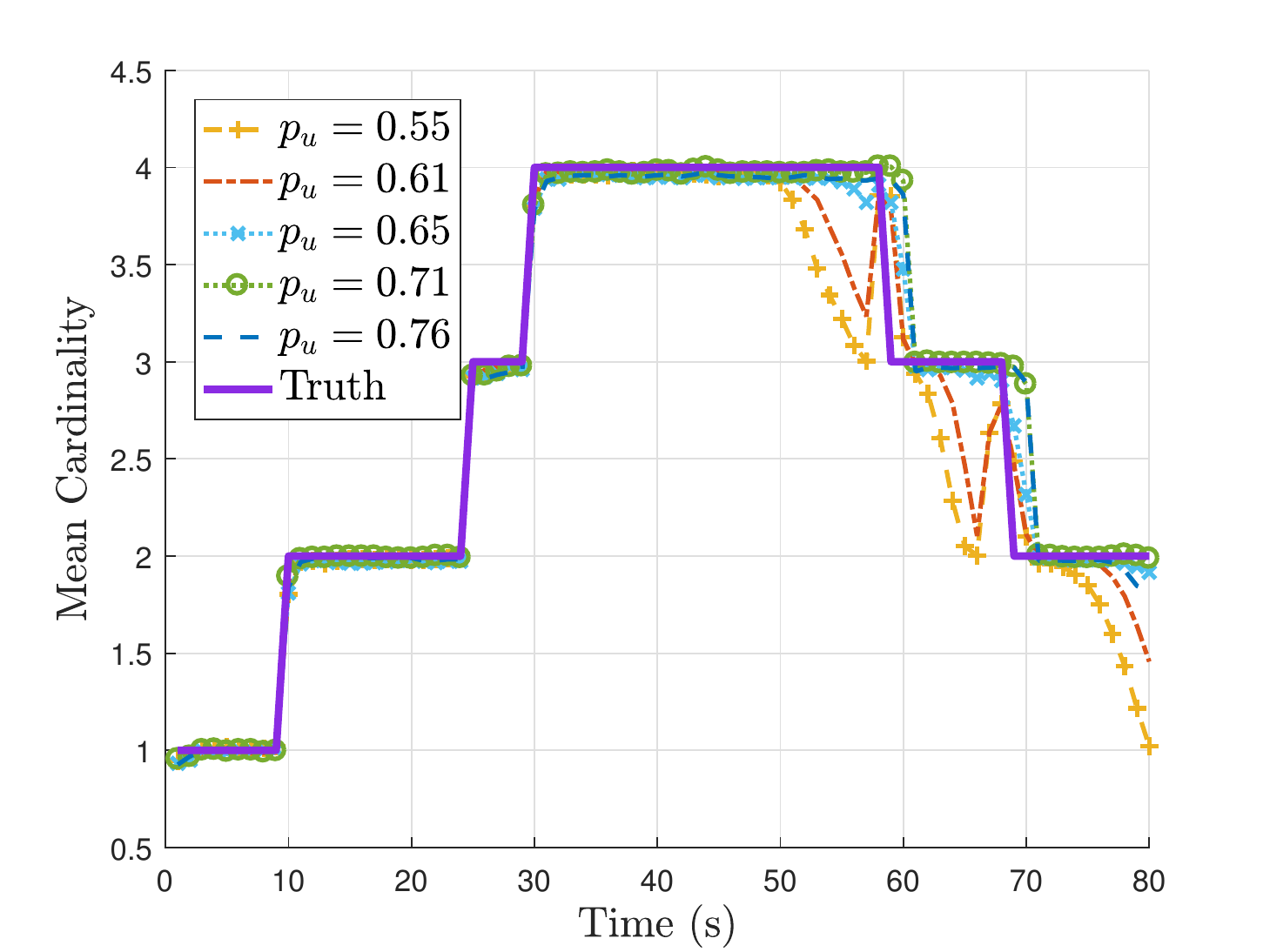}}
		\caption{Simulation results of OSPA with various $p_{u}$}\label{simulation2}
	\end{figure*}

\begin{figure*}
		\centering
		\subcaptionbox{\label{t1}}
		{\includegraphics[width=.45\linewidth]{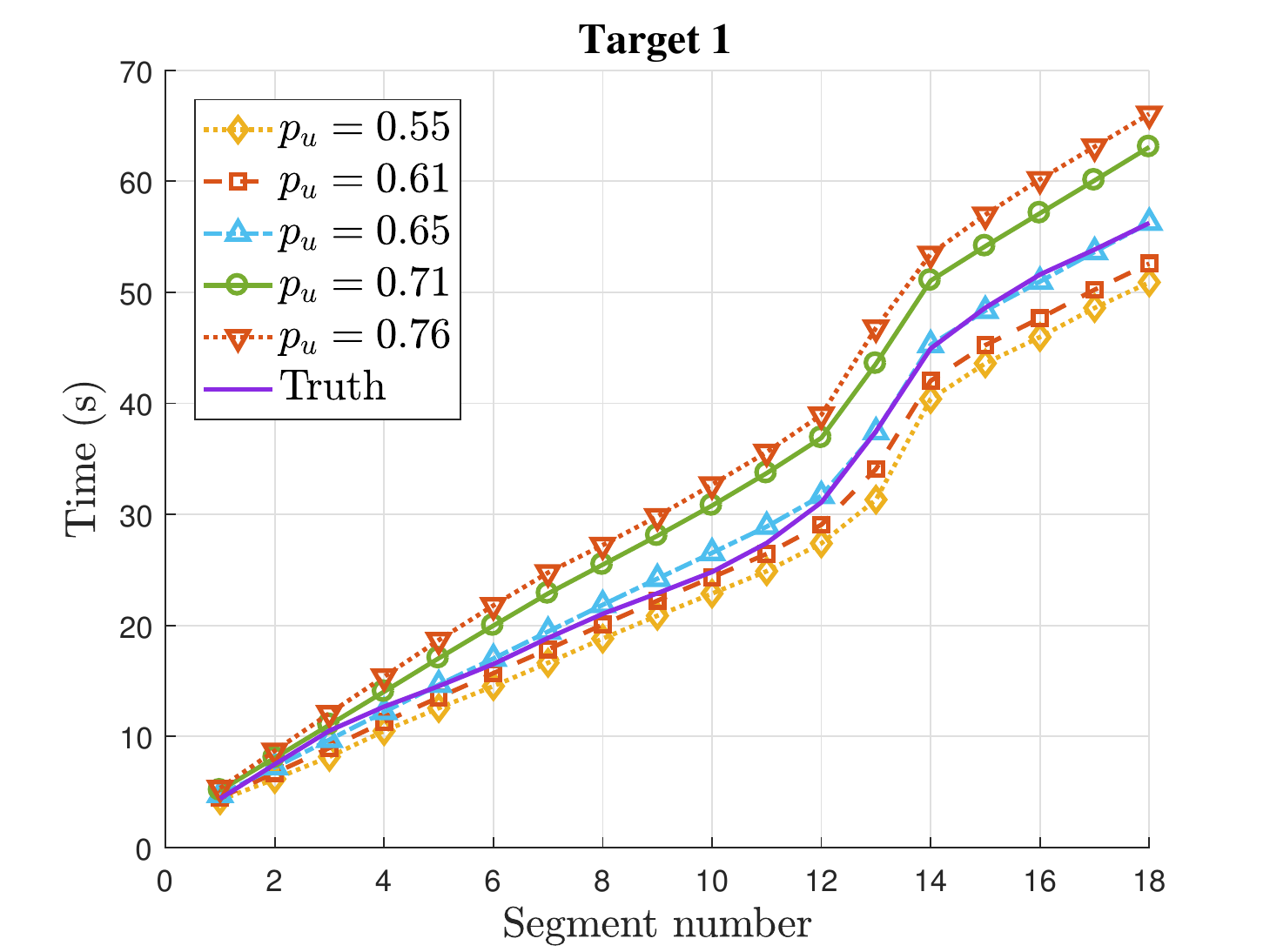}}
		\subcaptionbox{\label{t2}}
		{\includegraphics[width=.45\linewidth]{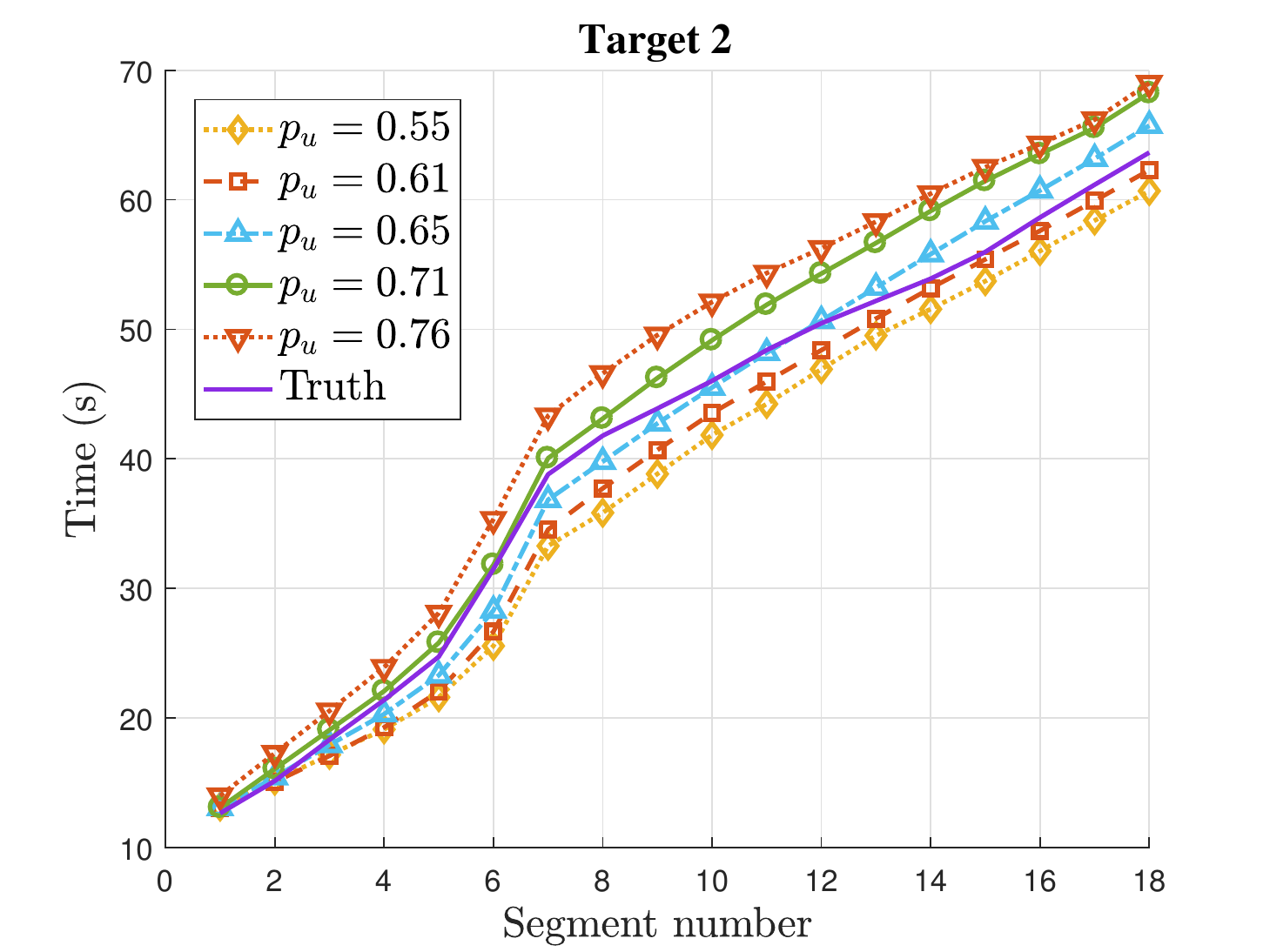}}
		\subcaptionbox{\label{t3}}
		{\includegraphics[width=.45\linewidth]{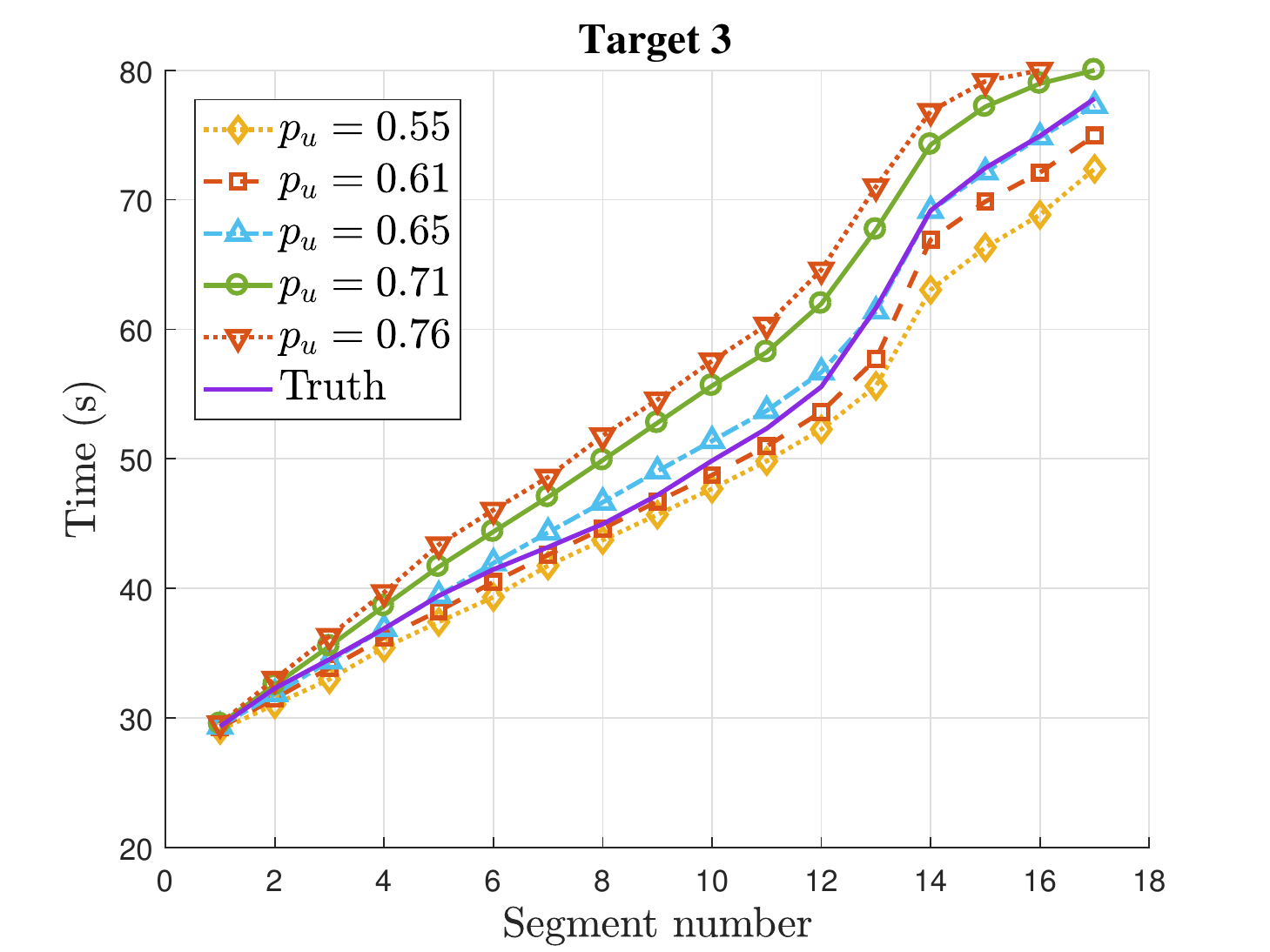}}
		\subcaptionbox{\label{t4}}
		{\includegraphics[width=.45\linewidth]{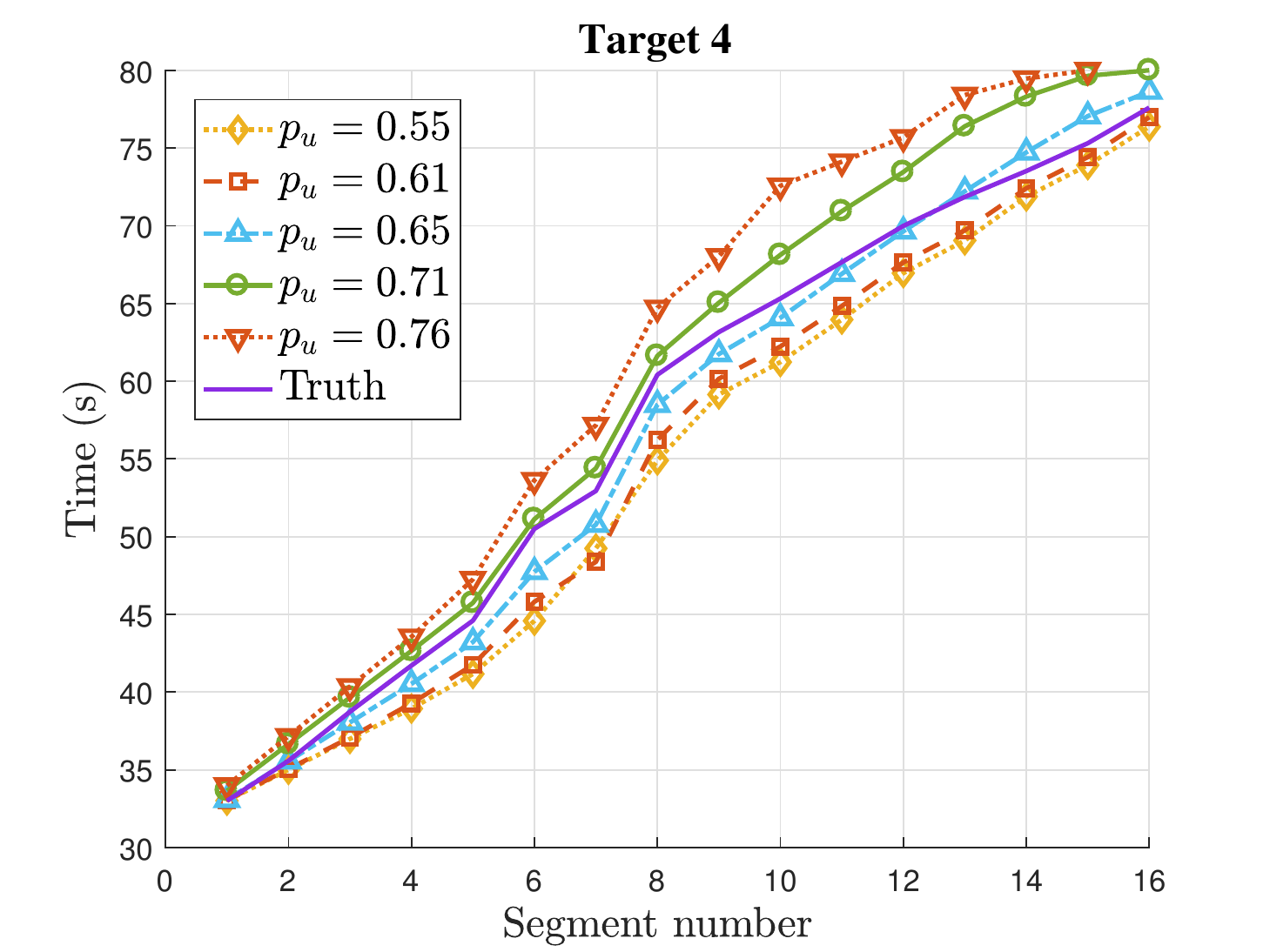}} 
		\caption{Simulation results of mean update time of different targets with various $p_u$} \label{mt}  
	\end{figure*}

	The simulation results of mean OSPA distance and mean cardinality are depicted in Fig. \ref{simulation2}, and the statistics of estimation errors are summarized in Table \ref{simulation2-tb}. From the recorded results, $p_u=0.65$ provides better estimation performance than the other four cases in terms of the OSPA distance. This can be attributed to the fact that the  constraint hypothesis is updated accurately and hence the road segment is correctly allocated to the moving vehicle, as shown in Fig. \ref{mt}. We can also note from Fig. \ref{mt} that the mean update time corresponding to $p_u=0.55$ and $0.61$ of all targets is smaller than the true update time, meaning the update is confirmed earlier than the real transformation, resulting in a sharp decrease in tracking accuracy, as shown in Fig. \ref{simulation2} and Table \ref{simulation2-tb}. This result reveals that the constraint hypothesis update is sensitive to the segment transformation and hence contributes to a premature confirmation of the upcoming segment if the value of $p_u$ is too small. Conversely, the constraint hypothesis update module is believed to be conservative when the exact segment transformation occurs with bigger value of $p_u$: the vehicles are still believed to move on the segments which deviate away from the correct ones. This can be confirmed by the existence of a time delay of mean update time corresponding to $p_u=0.71$ and $0.76$ compared with the real update time, as shown in Fig. \ref{mt}).       
	
	\begin{table*}[hbt!]
		\centering
		\caption{Statistics of estimation errors with various $p_{u}$}\label{simulation2-tb}
		\begin{tabular}{@{}ccccccc@{}}
			\toprule
			& \multicolumn{2}{c}{\textbf{OSPA distance}}      & \multicolumn{2}{c}{\textbf{Localization error}} & \multicolumn{2}{c}{\textbf{Cardinality error}}  \\ \cmidrule(l){2-7} 
			\multirow{-2}{*}{$p_{u}$} & Mean                          & Std    & Mean                          & Std    & Mean        & Std    \\ \midrule
			0.55                      & 5.9657       & 2.7674   & 4.1399  & 1.7025 & 1.8258      & 2.4970 \\
			0.61                      & {3.9526} & 1.9751 & 2.8203  & {1.1965} & {1.1323}  & 1.5708 \\
			0.65                      & \textbf{2.8730} & \textbf{1.0845} & \textbf{2.0862} & \textbf{0.4389} & {0.7869} & \textbf{0.9271} \\
			0.71                      & 3.8910    &1.9522  & 3.1341 &1.4810 & \textbf{0.7569}    &1.4530 \\
			0.76                      & 6.1254     &3.4708  &5.2177   & 3.1683    & 0.9077      &1.4202  \\ \bottomrule
		\end{tabular}
	\end{table*}
	
	\subsection{Performance of Modified Track Update}
\begin{figure*}[hbt!]
		\centering
		\subcaptionbox{Simulation results of mean OSPA distance\label{simulation3(1)-d}}
		{\includegraphics[width=.45\linewidth]{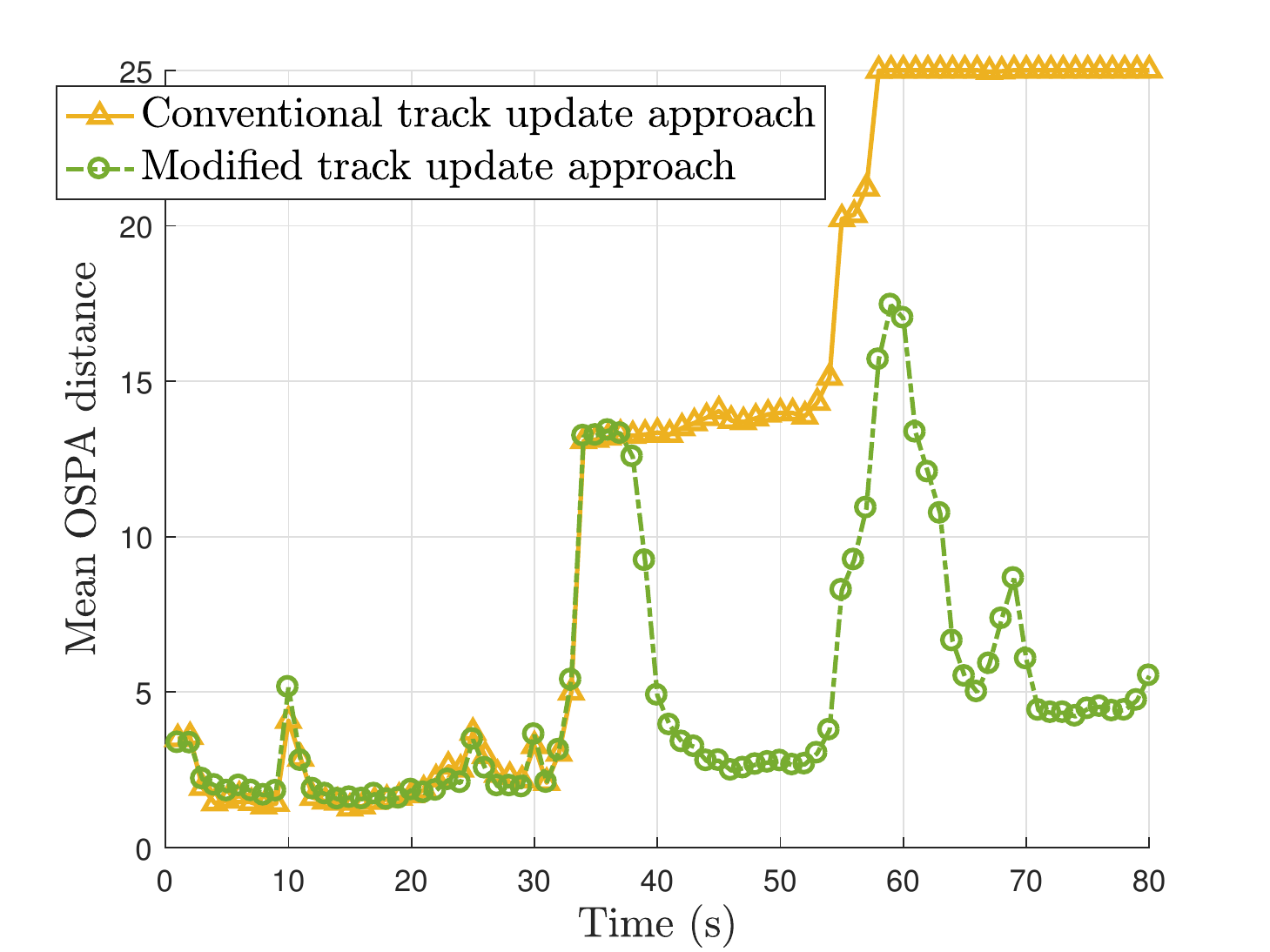}}
		\subcaptionbox{Simulation results of mean cardinality\label{simulation3(1)-c}}
		{\includegraphics[width=.45\linewidth]{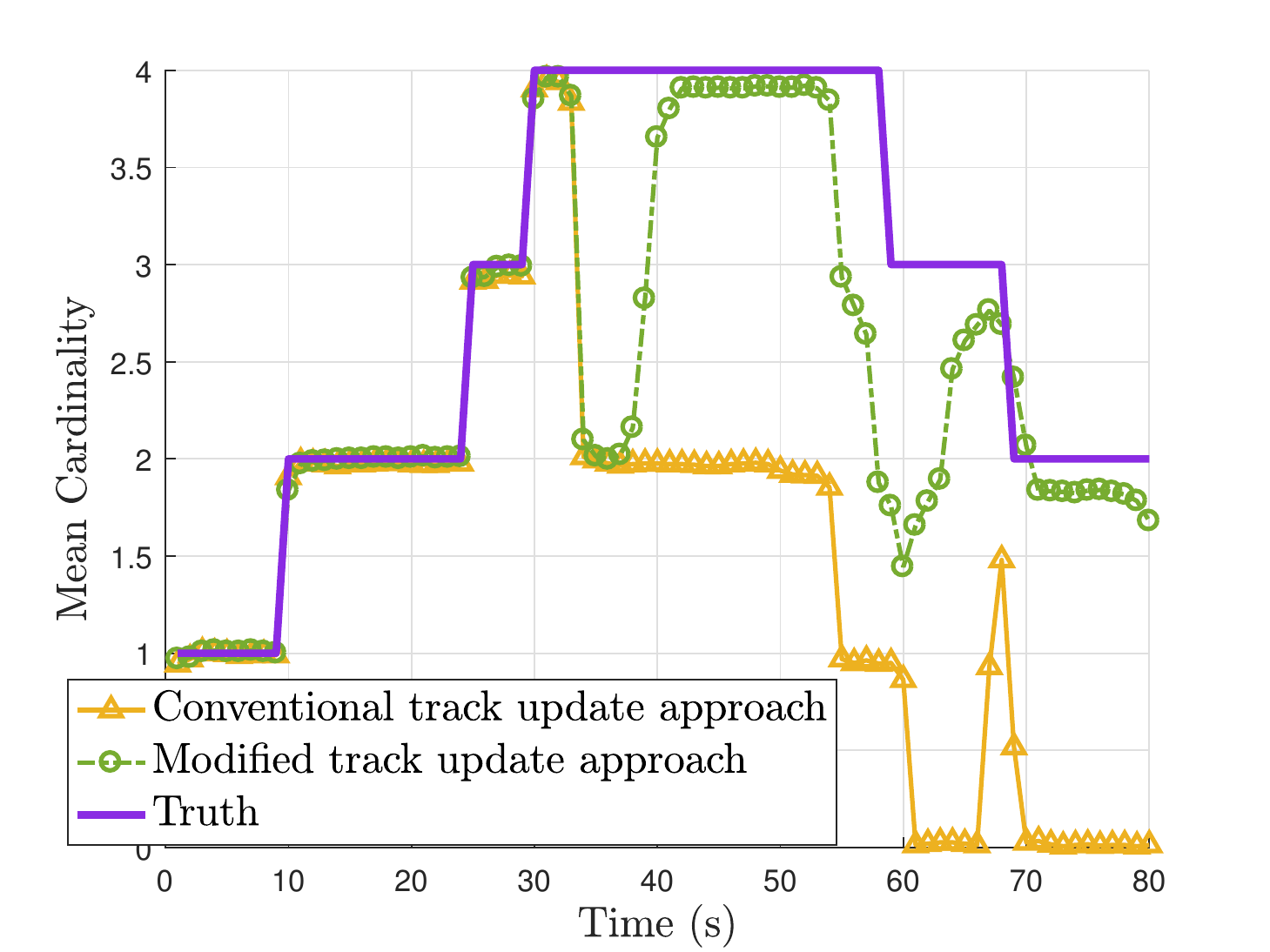}}
		\caption{OSPA comparison between conventional and modified track update approach.}\label{simulation3(1)}
	\end{figure*}

	In Section \ref{sec4}, we modified the track update approach to recognize a temporarily disappeared target. To verify the effectiveness of proposed method, we assume that the roads are partially blocked, ranging from $-2.126\times10^{2}$km to $-2.125\times10^{2}$km in $X$-direction. A vehicle returns no measurement once it enters into this blocked region. The simulation results of OSPA distances obtained by conventional track update logic and the proposed approach are depicted in Fig. \ref{simulation3(1)}. The results show that both methods are able to confirm new born tracks, i.e., the small peaks appear at around 1s, 10s, 25s and 30s shown in Fig. \ref{simulation3(1)-d}. However, only the algorithm incorporated with the modified track update approach can recognize reappeared targets as shown in Fig. \ref{simulation3(1)}. More specifically, it can be observed that some targets move into the blocked area at around 30s and 52s, and both methods believe the targets disappear. When the targets get through this blocked area and reappear at about 36s and 60s, the modified approach confirms these targets successfully while the conventional approach fails.
	
\begin{figure*}[hbt!]
		\centering
		\subcaptionbox{Simulation results of mean OSPA distance\label{simulation3(2)-d}}
		{\includegraphics[width=.45\linewidth]{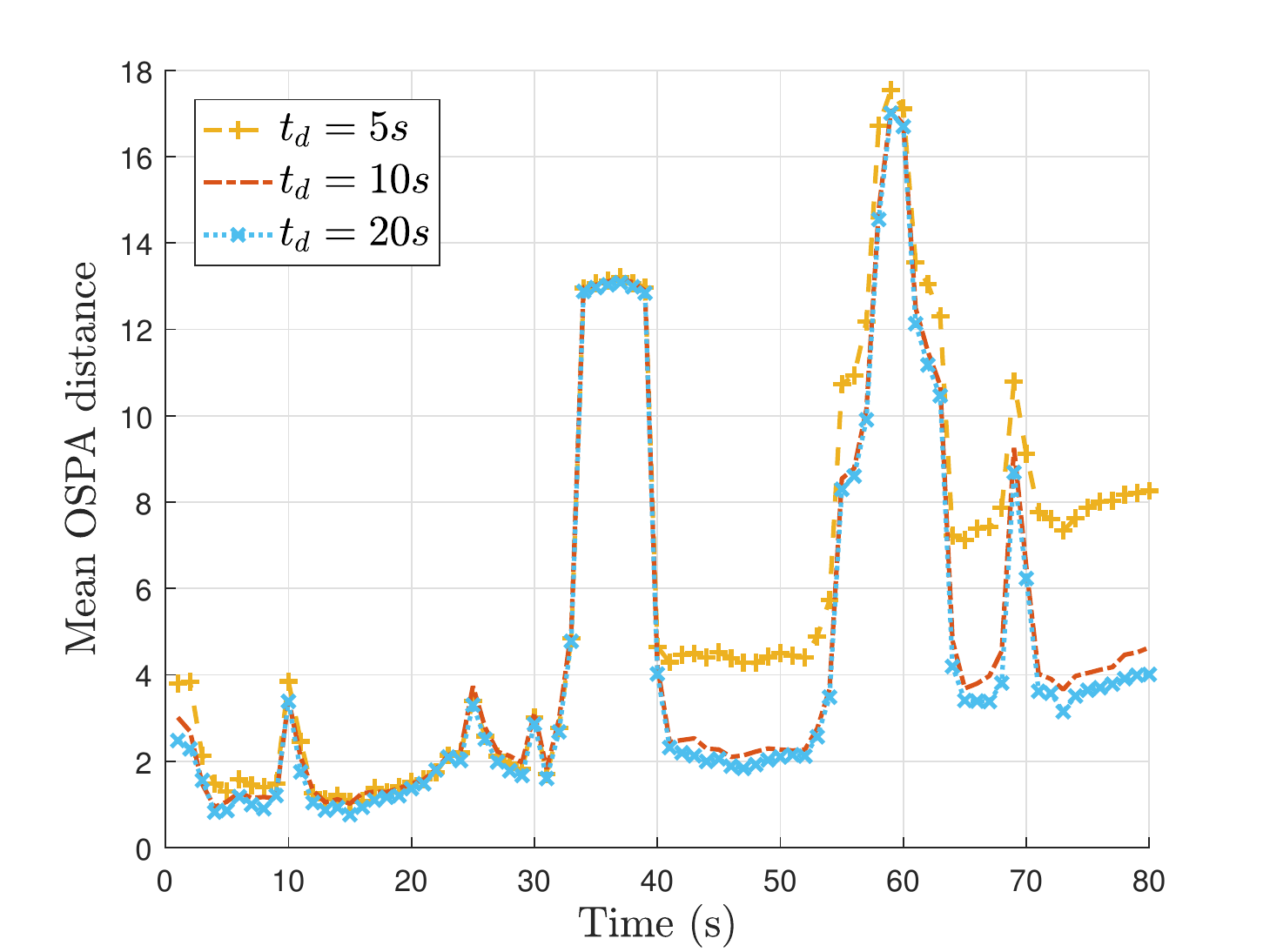}}
		\subcaptionbox{Simulation results of mean cardinality\label{simulation3(2)-c}}
		{\includegraphics[width=.45\linewidth]{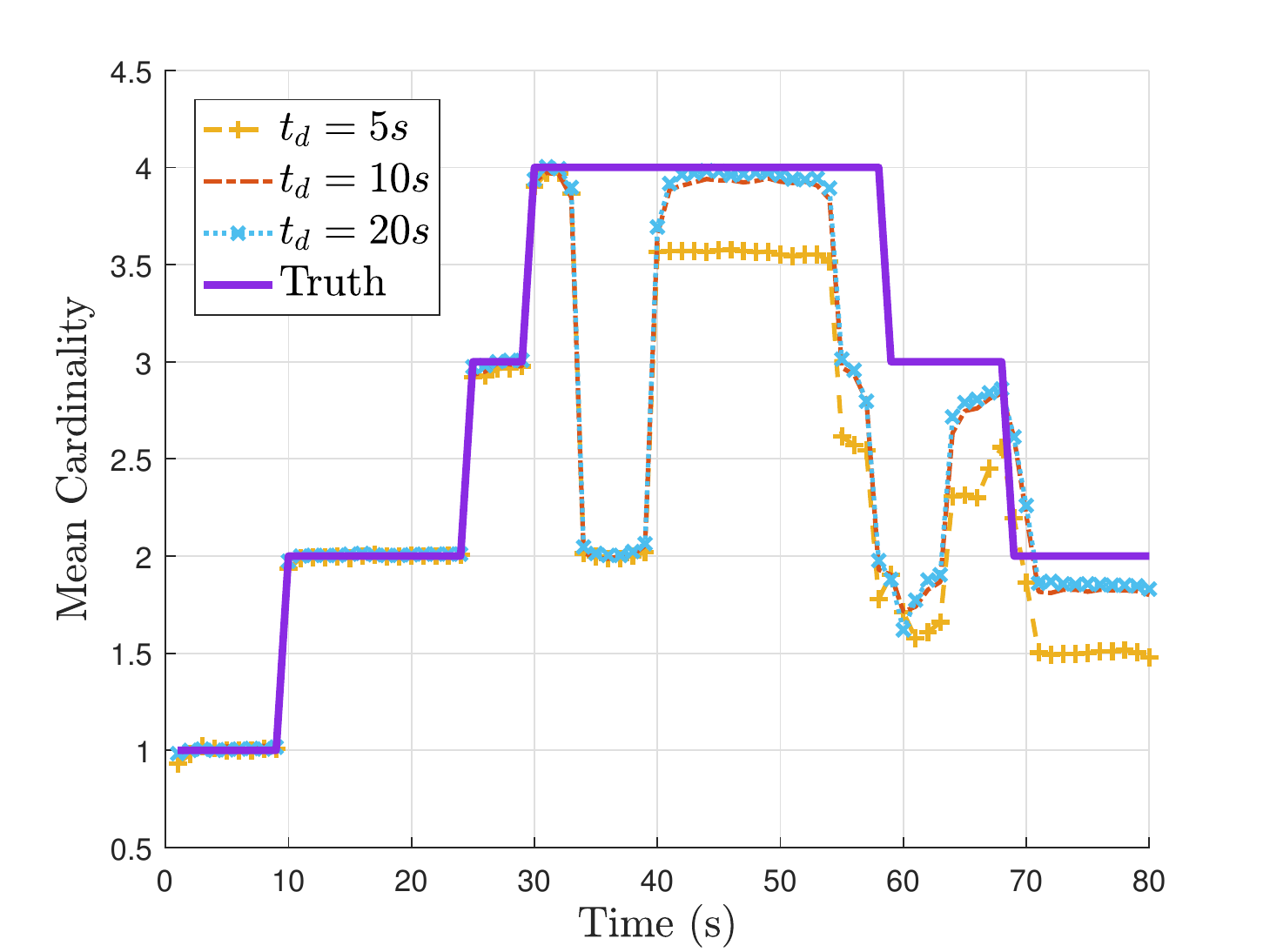}}
		\caption{Simulation results of OSPA with various $t_{d}$.}\label{simulation3(2)}
	\end{figure*}

	Now, let us check the performance of the proposed track update logic with various reconfirmation time thresholds $t_d$=5s, 10s, 20s. Figure \ref{simulation3(2)} presents the simulation results of OSPA, and Table \ref{simulation3(2)-tb} summarizes the statistics of the estimation errors. It is clear from this figure that when the value of $t_d$ is small, the proposed approach might fail to confirm the reappeared target since the temporarily disappeared target is removed before it reappears, as shown in Fig. \ref{simulation3(2)}. Even the tracking quality improves when the value of $t_d$ increases, extremely large $t_d$ might degrade the tracking performance. This can be attributed to the fact that the accumulation of tentative tracks provides high probability of false track confirmation.
	%\begin{figure}[hbt!]
		%	\centering
		%	\subcaptionbox{Simulation results of mean OSPA distance error\label{simulation3(3)-d}}
		%	{\includegraphics[width=.49\linewidth]{simulation2(3)-d.eps}}
		%	\subcaptionbox{Simulation results of mean cardinality\label{simulation3(3)-c}}
		%	{\includegraphics[width=.49\linewidth]{simulation2(3)-c.eps}}
		%	\caption{Simulation results of 300 Monte-Carlo runs corresponding to various $\mathbf{P}_{re}^{i}$}\label{simulation3(3)}
		%\end{figure}
	
	%\begin{table}[hbt!]
		%	\centering
		%	\caption{Simulation results summarized corresponding to conventional and modified track updating logic}\label{simulation3(1)-tb}
		%	\begin{tabular}{@{}ccccccc@{}}
			%		\toprule\toprule
			%		& \multicolumn{2}{c}{OSPA distance}      & \multicolumn{2}{c}{Localization error} & \multicolumn{2}{c}{Cardinality error} \\ \cmidrule(l){2-7} 
			%		\multirow{-2}{*}{Track updating logic} & Mean    & Std    & Mean   & Std    & Mean    & Std    \\ \midrule
			%		Conventional                           & 12.4383 & 9.6241 & 1.5939 & 2.1186 & 10.8444 & 9.5385 \\
			%		Modified & {4.4287} & 4.1908 & {1.7098} & 1.0245 & {2.7189} & 3.964 \\ \bottomrule\bottomrule
			%	\end{tabular}
		%\end{table}
	
	\begin{table*}[hbt!]
		\centering
		\caption{Statistics of estimation errors with various $t_d$ }\label{simulation3(2)-tb}
		\begin{tabular}{@{}ccccccc@{}}
			\toprule
			& \multicolumn{2}{c}{\text{OSPA distance}}      & \multicolumn{2}{c}{\text{Localization error}} & \multicolumn{2}{c}{\text{Cardinality error}}  \\ \cmidrule(l){2-7} 
			\multirow{-2}{*}{$t_d$} & Mean  & Std    & Mean                          & Std    & Mean   & Std    \\ \midrule
			5s  & 6.1553 & 4.3718 & \textbf{2.0503}  & \textbf{0.8700} & {4.1050 }  & 4.2167 \\
			10s & {5.3826}  &4.1917 &2.0959 & {0.8866} & {3.2867}  & 4.0689 \\
			20s & \textbf{5.2214} & \textbf{3.9012} &2.2087 & {0.9220} &\textbf{3.0127}   &\textbf{3.7132} \\ \bottomrule
		\end{tabular}
	\end{table*}
	%\begin{table}[hbt!]
		%%	\centering
		%	\caption{Simulation results summarized corresponding to various $\mathbf{P}_{re}^{i}$}\label{simulation3(3)-tb}
		%	\begin{tabular}{@{}ccccccc@{}}
			%		\toprule\toprule
			%		& \multicolumn{2}{c}{OSPA distance}     & \multicolumn{2}{c}{Localization error} & \multicolumn{2}{c}{Cardinality error}  \\ \cmidrule(l){2-7} 
			%		\multirow{-2}{*}{$\mathbf{P}_{re}^{i}$} & Mean                         & Std    & Mean               & Std               & Mean                          & Std    \\ \midrule
			%		$\mathbf{P}_a$ &
			%		5.2644 &
			%		\multicolumn{1}{l}{4.412} &
			%		\multicolumn{1}{l}{{ 3.9642}} &
			%		\multicolumn{1}{l}{1.1248} &
			%		3.5762 &
			%		4.1464 \\
			%		$\mathbf{P}_b$ &
			%		{ 4.1582} &
			%		4.2129 &
			%		{ 1.7631} &
			%		1.0768 &
			%		{ 2.3951} &
			%		3.8153 \\
			%		$\mathbf{P}_c$                          & { 5.751} & 4.0328 & 2.1697             & 1.3461            & { 3.5812} & 4.1549 \\ \bottomrule\bottomrule
			%	\end{tabular}
		%%\end{table}
	
	\subsection{Performance with Different Types of Domain Knowledge}
	The performance improvement by incorporating multiple sources of domain knowledge into the target tracking system is verified in this subsection. For this purpose, the proposed method integrated with four different constraint combinations is considered, i.e., \emph{Condition 1}: no constraints are included; \emph{Condition 2}: only velocity direction constraint is considered (corresponds to Case 1 in Table \ref{conscmbn}); \emph{Condition 3}: velocity direction and position constraints are considered (corresponds to Case 4 in Table \ref{conscmbn}); and \emph{Condition 4}: velocity direction, position and speed limit are constrained (corresponds to Case 7 in Table \ref{conscmbn}). The simulations are carried out with two different scenarios: unblocked and partially blocked. The blocked region remains the same as the previous subsection.
	
\begin{figure*}[hbt!]
		\centering
		\subcaptionbox{Simulation results of mean OSPA distance\label{simulation4-noblock-d}}
		{\includegraphics[width=.45\linewidth]{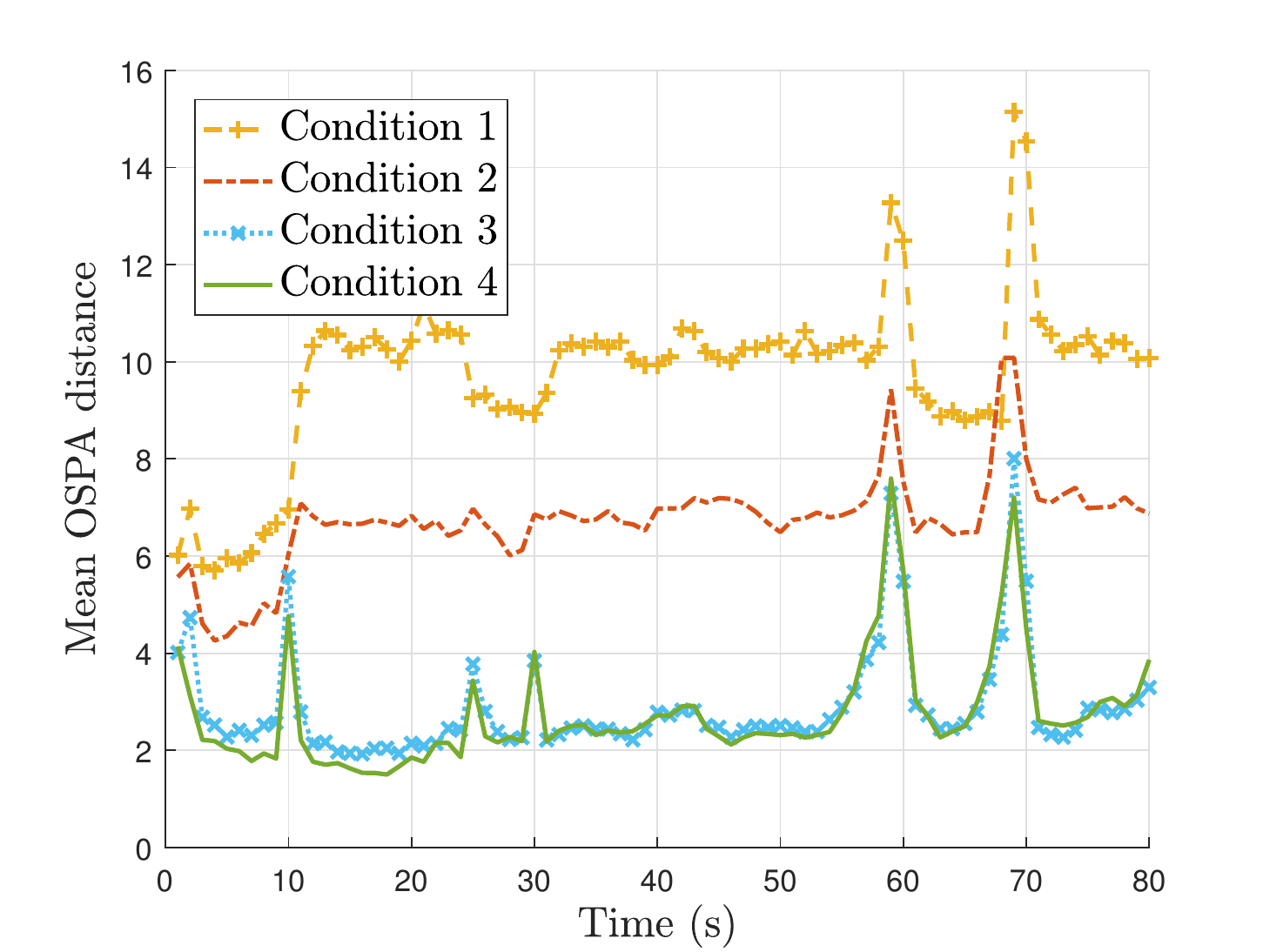}}
		\subcaptionbox{Simulation results of mean cardinality\label{simulation4-noblock-c}}
		{\includegraphics[width=.45\linewidth]{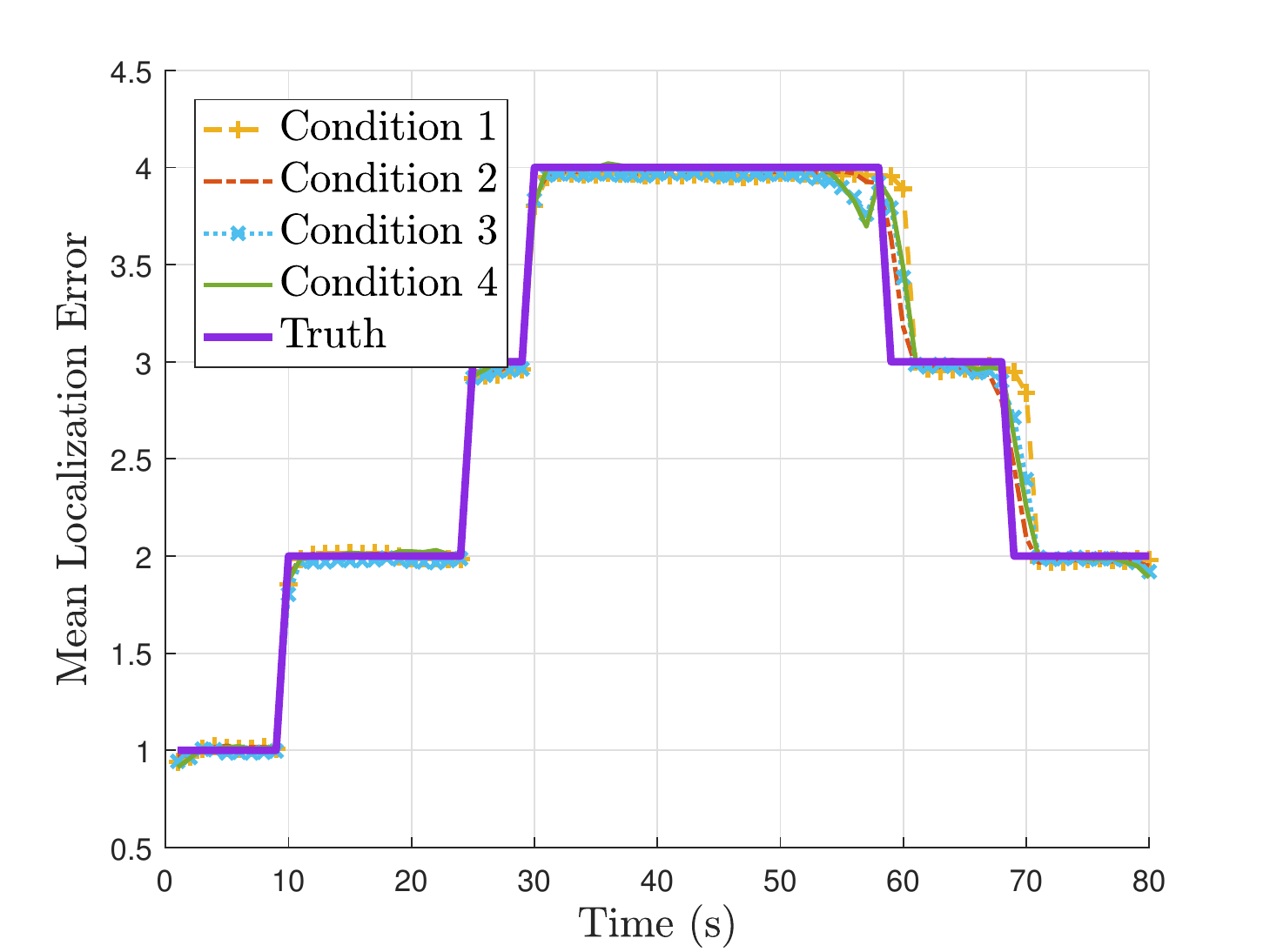}}
		\caption{Simulation results of OSPA with various conditions in the unoblocked scenario.}\label{simulation4-noblock}
	\end{figure*}

\begin{figure*}[hbt!]
		\centering
		\subcaptionbox{Simulation results of mean localization error\label{simulation4-block-d}}
		{\includegraphics[width=.45\linewidth]{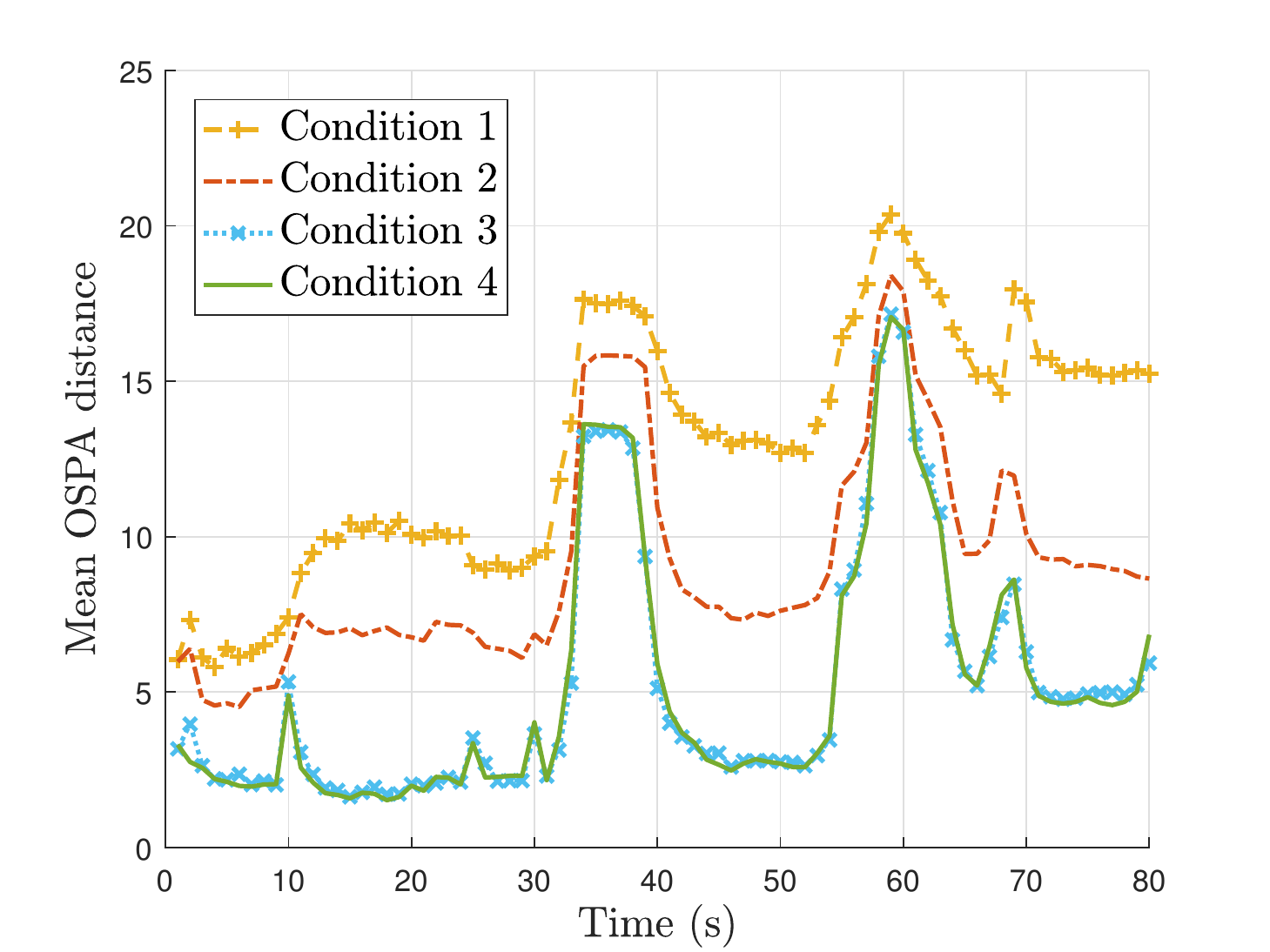}}
		\subcaptionbox{Simulation results of mean cardinality\label{simulation4-block-c}}
		{\includegraphics[width=.45\linewidth]{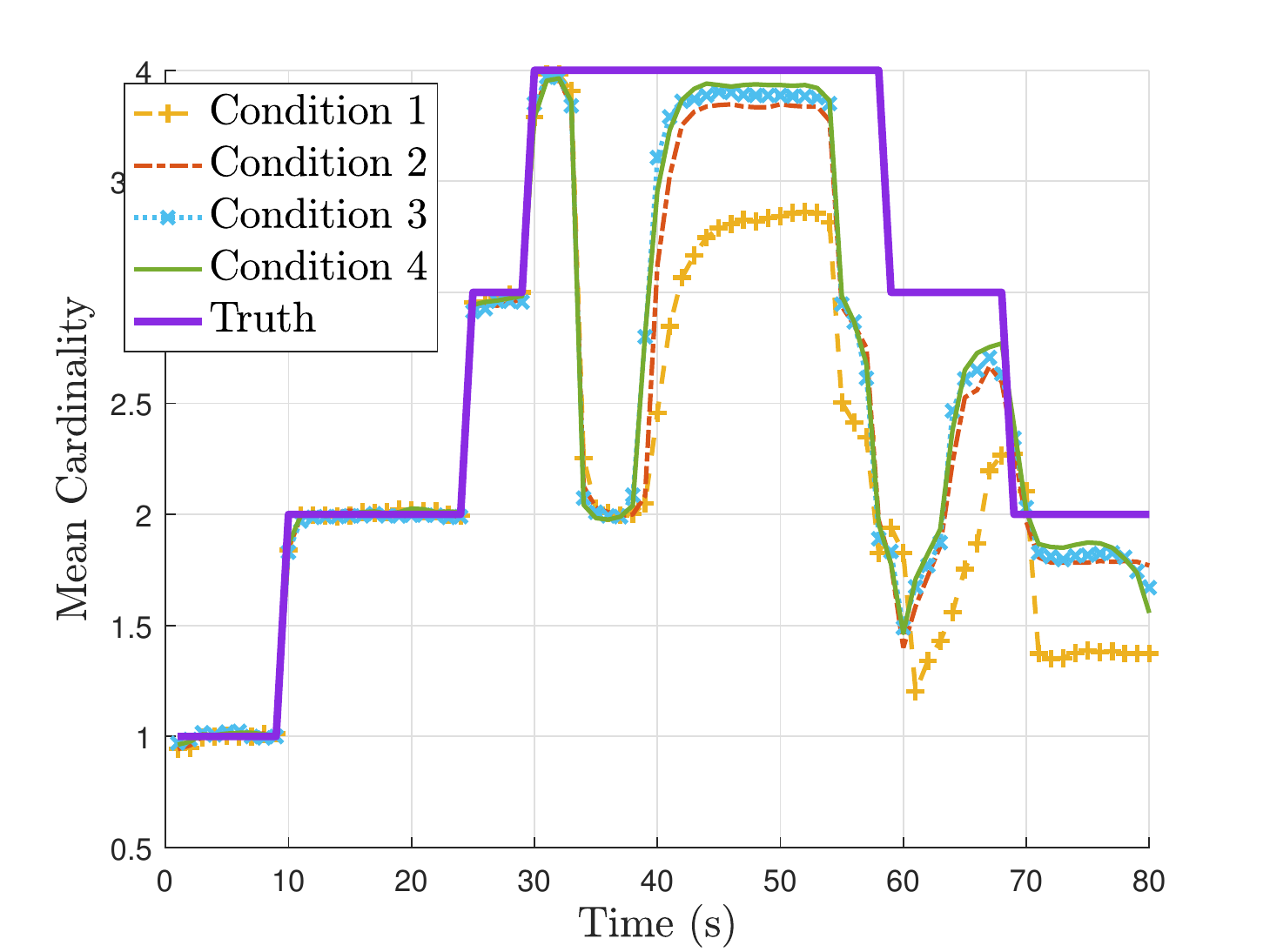}}
		\caption{Simulation results of OSPA with various conditions in the  partially blocked scenario.}\label{simulation4-block}
	\end{figure*}

	Simulation results of OSPA distance and mean cardinality of the two different scenarios are shown in Figs. \ref{simulation4-noblock} and \ref{simulation4-block}, respectively. The statistics of the estimation errors are summarized in Table \ref{simulation4-tb} for clarity. According to the simulation results, it is clear that the tracking performance is improved by incorporating multiple sources of domain knowledge. From Figs. \ref{simulation4-noblock} and \ref{simulation4-block}, we can also observe that the estimation errors decrease with the more valuable constraint information. Interestingly, leveraging domain knowledge as state constraint in JPDA is helpful to recognize temporarily disappeared targets.
	\begin{table}[hbt!]
		\centering
		\caption{Statistics of estimation errors with various conditions}\label{simulation4-tb}
		\begin{tabular}{@{}cccccccc@{}}
			\toprule
			&
			&
			\multicolumn{2}{c}{\text{OSPA distance}} &
			\multicolumn{2}{c}{\text{Localization error}} &
			\multicolumn{2}{c}{\text{Cardinality error}} \\ \cmidrule(l){3-8} 
			\multirow{-2}{*}{\text{Scenario}} & \multirow{-2}{*}{\text{Condition}} & Mean    & Std                           & Mean   & Std                           & Mean   & Std    \\ \midrule
			& 1   & 9.7514  & {1.6891} & {8.9391} & 1.5065 &0.8123  & 1.3769 \\
			& {2} &6.7358  &\textbf{0.9614} &{6.1818} &0.7824 &\textbf{0.5540} &\textbf{0.6945} \\
			& {3}    &2.8749 & {1.0975} &\textbf{2.0715}   &\textbf{0.4290}  &0.8035 &0.9493 \\
			\multirow{-4}{*}{Unblocked} &
			4 &	\textbf{2.7450} &1.1335 &2.1825 &	{0.4804}&	{0.5626} &0.9338 \\ \midrule
			& 1                           & 12.9457 & 3.9260 & 7.8959   & 2.0369  & 5.0498 & 4.5606 \\
			& 2                           &9.0860    & 3.4128  & 5.7228  & 1.2270  &3.3632   & 3.9734 \\
			& 3                           &5.2114 & \textbf{4.0260}  & \textbf{2.1130} & \textbf{0.8868}   & 3.0985   &\textbf{3.8746}  \\
			\multirow{-4}{*}{Partially blocked} &
			4 &	\textbf{5.1736} & 4.0534 &2.2117  &{0.9422} &\textbf{2.9619}&3.9301 \\ \bottomrule
		\end{tabular}
	\end{table}
	
	%%%%%%%%%%%%%%%%%%%%%%%%%%%%%%%%%
	\section{Conclusion}\label{sec7}
	In this paper, a novel multiple ground vehicle tracking method integrated with domain knowledge is proposed for automatic aerial surveillance by a UAV. The topography information of the surveillance region is extracted from the local map as domain knowledge. This information is incorporated into the JPDA filter as state constraints to improve the tracking performance. A modified track update method is also developed for the purpose of reconfirming temporarily disappeared targets. In order to properly utilize the domain knowledge, the VS-MM concept is leveraged to allocate a target to a specific road segment. The numerical simulation results show the proposed algorithm provides reliable tracking of ground moving targets and the performance can be improved with a proper choice of the required tuning parameters.

	
%	\bibliographystyle{apsrev4-1}
	\bibliography{sn-bibliography}
	\bibliographystyle{abbrv}
%	\input 
\end{document}